\documentclass[aps,prd,twocolumn,showpacs,superscriptaddress,nofootinbib,a4paper,10pt, reprint,eqsecnum]{revtex4-2}

\usepackage{newtxmath,newtxtext}
\usepackage{mathtools}
\usepackage{bm}
\usepackage{graphicx}
\usepackage{dcolumn}
\usepackage[dvipsnames]{xcolor}
\usepackage{hyperref}
\usepackage{booktabs}
\usepackage{multirow}
\usepackage{makecell}

\hypersetup{colorlinks=true,linkcolor=blue,citecolor=ForestGreen,urlcolor=blue}
\usepackage{orcidlink}

\newcommand{\phidot}{\dot{\phi}}
\newcommand{\phidd}{\ddot{\phi}}
\newcommand{\Hdot}{\dot{H}}
\newcommand{\EGB}{\mathcal{G}}
\newcommand{\rDE}{\rho_{\rm \phi}}

\newcommand{\weff}{w_{\rm eff}}
\newcommand{\Ode}{\Omega_{\phi}}
\newcommand{\Om}{\Omega_{\rm m}}
\newcommand{\Or}{\Omega_{\rm r}}
\newcommand{\Og}{\Omega_{\rm GB}}

\newcommand{\pp}{\partial}

\newcommand{\ct}{\ensuremath{c_{\rm T}^{2}}}
\newcommand{\cs}{\ensuremath{c_{\rm s}^{2}}}

\newcommand{\nb}{\ensuremath{ \nabla }}
\newcommand{\m}{\ensuremath{{\mu \nu}}}

\newcommand{\mph}{\ensuremath{m_\phi^2}}
\newcommand{\rr}{\ensuremath{\rho_{\rm r}}}
\newcommand{\rmm}{\ensuremath{\rho_{\rm m}}}

\newcommand{\eos}{\ensuremath{w_{\rm eff}}}

\begin{document}

	\title{Oscillatory dark energy with phantom crossings in Einstein--Gauss--Bonnet gravity}
	
	\author{Saddam Hussain\orcidlink{0000-0001-6173-6140}}
	\email{saddamh@zjut.edu.cn}
	
	\affiliation{Institute for Theoretical Physics and Cosmology, Zhejiang University of Technology,
		Hangzhou 310023, China}
	
	\author{Sandip Biswas\orcidlink{0009-0002-4040-8791}}
	\email{sandipb20@iitk.ac.in}
	\affiliation{Department of Physics, Indian Institute of Technology, Kanpur, Kanpur-208016, Uttar Pradesh, India}
	
	\date{\today}
	
	\begin{abstract}
		Recent baryon acoustic oscillation measurements from the Dark Energy
		Spectroscopic Instrument (DESI), combined with supernova observations, show a mild preference for an evolving
		dark-energy equation of state, in which dark energy remains phantom-like at
		higher redshift while a transition toward the quintessence regime emerges
		around $z\lesssim0.5$. Beyond the monotonic evolution of the equation of
		state, an oscillatory dark-energy scenario provides a potential alternative
		realization of dynamical dark energy. We study such an oscillatory scenario
		within the Einstein-scalar-Gauss--Bonnet gravity framework, where a scalar
		field with a hybrid exponential-quadratic potential couples nonminimally to
		the Gauss--Bonnet (GB) invariant. We choose a Gaussian GB coupling function
		localized around the minimum of the potential. During the late-time epoch,
		the field thereby oscillates around the potential minimum and eventually
		exits toward a stable de Sitter attractor phase. In the minimally coupled
		limit, the effective equation of state oscillates in the quintessence
		regime at low redshift, $z\sim0$, without crossing the phantom divide. The
		nonminimal coupling amplifies the oscillations and drives the effective
		equation of state across the phantom divide multiple times. Although a
		sufficiently strong coupling can induce negative scalar and tensor sound
		speeds, rendering the perturbation modes unstable. We identify a region of
		parameter space in which the model undergoes multiple phantom crossings
		while remaining free of ghost and gradient instabilities and approaching a
		stable de Sitter attractor in the future.
	\end{abstract}
	
	\maketitle
	
	\section{Introduction}
 
 Technological advances in the ${\rm 21^{st}}$ century have enabled us to probe the Universe with unprecedented precision, allowing us to reconstruct its cosmic history while revealing increasingly challenging and counter-intuitive aspects of its nature. One of the most intriguing discoveries is that an unknown form of matter, which gravitates but does not interact electromagnetically with ordinary baryonic matter, is one of the dominant components of the Universe and plays a crucial role in the formation of the large-scale structures we observe today, contributing nearly $25\%$ of the total cosmic energy budget. Ordinary baryonic matter constitutes roughly $5\%$ and radiation $10^{-3}\%$ of this budget, interacts electromagnetically, and {makes up the familiar matter that can be detected in laboratory experiments}. The remaining energy budget is attributed to an unknown component responsible for driving the accelerated expansion of the Universe, whose origin and fundamental nature remain unclear. These unknown components have significantly challenged our understanding of fundamental physics across both microscopic and macroscopic scales.
 
 To explain these entities, {the simplest cosmological framework is built by extending General Relativity with} a cosmological constant, $\Lambda$, that drives cosmic acceleration, together with a pressureless, non-relativistic fluid responsible for structure formation. Together, these components constitute the $\Lambda$CDM framework, (where CDM refers to cold dark matter), which provides an excellent fit to the majority of cosmological observations \cite{SupernovaCosmologyProject:1998vns,SupernovaSearchTeam:1998fmf,WMAP:2003elm,Sherwin:2011gv,Wright:2007vr,DES:2016qvw,DES:2021esc,SDSS:2005xqv,Primack:1997av,DelPopolo:2007dna,Diao:2023tor,Mina:2020eik,Blumenthal:1984bp,Baccigalupi:2006ww,Planck:2015fie,Planck:2018vyg,Eisenstein:2005sbt,Percival:2007yw,BOSS:2016wmc,Li:2008vf,DES:2017tss,LSST:2008ijt}. Nevertheless, the model suffers from several {fundamental} problems. For instance, the observed value of $\Lambda$ differs by up to $120$ orders of magnitude from the value expected from quantum field theory, leading to the well-known fine-tuning problem. The model also fails to explain the cosmic coincidence problem, which concerns why dark energy comes to dominate the energy budget only near the present epoch rather than much earlier or later \cite{Copeland:2006wr,Weinberg:1988cp,Rugh:2000ji,Padmanabhan:2002ji,Carroll:1991mt,Bengochea:2019daa,Kohri:2016lsj,Lopez-Corredoira:2017rqn}. Various theoretical models have therefore been proposed to address these issues by introducing physically motivated dynamical fields that mimic the behavior of the cosmological constant, including the quintessence field \cite{Peebles:2002gy,Hussain:2023kwk,Roy:2022fif}, $k$-essence scalar fields \cite{Armendariz-Picon:2000nqq,Armendariz-Picon:2000ulo,Chiba:1999ka,Armendariz-Picon:2005oog,Arkani-Hamed:2003pdi,Scherrer:2004au,Hussain:2024qrd}, and tachyonic fields \cite{Bagla:2002yn,Khoeini-Moghaddam:2018znw,Hussain:2022dhp}. Extensions of these models have also been considered in which the fields interact directly with dark matter through coupling functions \cite{Bertolami:2012xn,He:2011qn,Chatterjee:2021ijw,vandeBruck:2015ida,Kase:2020hst,Teixeira:2019hil,Bhattacharya:2022wzu,Hussain:2022osn,vandeBruck:2016jgg,Boehmer:2015kta,Boehmer:2015sha} or through modified energy-momentum relation \cite{Amendola:1999er,Boehmer:2008av,Bolotin:2013jpa,Wang:2016lxa,Montani:2024pou,Pan:2023mie,Hussain:2025uye}. Other approaches modify the gravitational sector by promoting $R\to f(R)$, where $f(R)$ is a generalized function of the Ricci scalar, thereby attempting to explain cosmic acceleration as a property of geometry rather than through the introduction of an additional matter component. However, such approaches face stringent theoretical and observational constraints, motivating scalar-tensor theories that couple a scalar field to curvature, such as $f(R,\phi)$ gravity \cite{Sotiriou:2008rp,DeFelice:2010aj,Guo:2013swa,Chatterjee:2024sjm,Hu:2007nk,Myrzakulov:2025jpk,Wang:2010mw}. Subsequently, these ideas culminated in the Horndeski framework, the most general scalar-tensor theory that exhibits second-order field equations and remains free from Ostrogradsky instabilities \cite{Horndeski:1974wa,Deffayet:2011gz,Kobayashi:2011nu,Kobayashi:2019hrl,Bellini:2014fua,Takahashi:2022mew,Fier:2025huc}. {Alternative formulations include modifications of gravity based on torsion rather than curvature, {giving rise to} teleparallel gravity and its extensions such as $f(T)$ gravity \cite{doi:https://doi.org/10.1002/3527608958.ch36,Cai:2015emx}. More recently, gravity has also been formulated in a flat, torsion-free geometry characterized by non-metricity, {resulting in} $f(Q)$ gravity \cite{Nester:1998mp,Heisenberg:2023lru}. These geometrical formulations are equivalent to General Relativity at the level of the gravitational action up to boundary terms, while providing alternative avenues for constructing modified-gravity theories and exploring deviations from the standard $\Lambda$CDM framework.

Besides the theoretical drawbacks of the concordance model, recent measurements of Baryon Acoustic Oscillations (BAO) from the Dark Energy Spectroscopic Instrument (DESI) have {revived} interest in dynamical dark energy, with the two-parameter CPL model showing a $2.6$--$3.9\sigma$ preference over the cosmological constant \cite{DESI:2024mwx,DESI:2025zgx,DES:2024upw,DES:2025sig}. When constrained by {observational} data, the CPL parameters {produce} a phantom-like behavior, with $w_{\rm de}<-1$ at high redshifts, $z>0.5$, before transitioning to the quintessence regime, $w_{\rm de}>-1$, at $z\lesssim0.5$. Such phantom dynamics at higher redshifts are, however, model dependent, as the CPL parametrization is one of the simplest extensions of the cosmological constant and is constructed as a Taylor expansion around the present value of scale factor $a=1$ \cite{Chevallier:2000qy,Linder:2002et}. The parametrization was initially proposed to capture the behavior of {quintessence} fields and to distinguish the signatures of different dark-energy models, including $k$-essence and modified-gravity models, while providing a phenomenological and model-independent description. {However, the future extrapolation $a>>1$ of the CPL parametrization leads to a divergent equation of state and therefore does not provide a consistent description of the late-time evolution of the Universe.} Therefore, one must either adopt another parametrization or invoke a field-theoretic prescription that reproduces CPL-like behavior in the past while providing a consistent extrapolation into the future.

Although phantom crossing is observationally viable, from a field-theoretic perspective it {requires} a negative kinetic term, which {renders the energy unbounded from below and consequently introduces} a ghost degree of freedom \cite{Nojiri:2026uvn}. From the particle-physics perspective, such ghost fields are generally considered theoretically problematic and cannot be consistently realized within a conventional, physically viable field-theoretic framework. A combination of two fields with positive- and negative-kinetic terms and appropriately chosen potentials, known as a quintom model \cite{Feng:2004ad}, can produce similar dynamics, allowing the dark-energy equation of state to cross the phantom divide, $w<-1$, and enter the quintessence regime, $w>-1$ \cite{Yang:2024kdo,Cai:2025mas}. {Yet} such models also suffer from the presence of ghost degrees of freedom and face challenges in achieving a consistent fundamental realization \cite{Cai:2009zp}. Similar behavior can also be realized within scalar-tensor theories through nonminimal couplings \cite{Wolf:2025jed,Ye:2024ywg}; {such couplings may, however, induce} variations in the effective gravitational constant on cosmological scales or generate an additional fifth force.

The particular signature of phantom behavior at higher redshifts is very likely to be a model artifact, it is important to investigate whether this behavior persists under more general parametrizations. The CPL parametrization can only exhibit monotonic evolution and does not provide sufficient degrees of freedom to capture additional variations in the equation of state. Thus, various parametric polynomial extensions of the equation of state have been proposed to capture additional degrees of freedom in the DESI BAO observational samples \cite{Escamilla:2021uoj,DESI:2025wyn,Ormondroyd:2025exu,Ormondroyd:2025iaf,Hussain:2025nqy}, which consistently hint at dynamics involving multiple crossings of the phantom divide, $w=-1$; {in other words, they suggest} an oscillatory equation of state for dark energy. Various phenomenological models with oscillatory parametrizations have been constructed using combinations of trigonometric functions \cite{Rezaei:2024vtg,Kessler:2025kju,Escamilla:2024fzq}. For instance, in Model~I of Ref.~\cite{Rezaei:2024vtg}, the oscillations occur predominantly in the distant past, with a sufficiently low frequency, and the corresponding oscillatory behavior decays as the system evolves toward lower redshifts. Similar dynamics can be seen in the remaining models.
 Likewise, the models in Ref.~\cite{Kessler:2025kju} exhibit similar dynamics, where the corresponding frequency for all the considered datasets becomes extremely small due to the particular choice of the functional form. These models, to some extent, behave similarly to the CPL parametrization, exhibiting phantom behavior at higher redshifts. This suggests that exploring oscillatory models within a generalized framework may reveal distinct signatures compared with fixed functional forms. Later, Ref.~\cite{Hussain:2026srf} addressed the shortcomings of these oscillatory frameworks by adopting a dynamical principle governing the full oscillatory behavior, in which a damped harmonic oscillator equation of state was introduced in the form of a second-order differential equation with a damping term that can counteract the oscillations. The dynamical nature of the model allows it to capture oscillatory phenomena either in the past or at lower redshifts. The model exhibits underdamped behavior at low redshifts depending on the supernova compilation, while {the equation of state} decays toward the equilibrium value $-1$ at higher redshifts. Due to the underdamped behavior, the equation of state crosses the phantom divide multiple times at lower redshifts, in contrast to the CPL parametrization \cite{Hussain:2026wss}.

In this paper, we realize the low-redshift phantom crossing within the framework of Einstein-scalar-Gauss-Bonnet (EsGB) gravity, in which the Gauss-Bonnet (GB) invariant non-minimally couples to the scalar field and is well motivated by string/M-theory
 \cite{Gross:1986mw,Bento:1995qc,Ferrara:1996hh,Antoniadis:1997eg,Nojiri:2005vv,Tsujikawa:2006ph,DeFelice:2006pg,Minamitsuji:2024twp,Tsujikawa:2022aar,DeFelice:2009rw,DeFelice:2006pg,Gross:1986mw,Bento:1995qc,Ferrara:1996hh}. The idea of oscillatory dark-energy dynamics from a field-theoretic perspective is not new. Earlier works have investigated oscillatory scenarios to address the cosmic coincidence problem \cite{Linder:2005dw}, periodic acceleration \cite{Rubano:2003er,Zhao:2006mn,Feng:2004ff}, structure formation, and the suppression of the low-multipole ISW effect \cite{Das:2013sca,Pace:2011kb}. In the field-theoretic approach, to realize an oscillatory scenario, one can consider a periodic or quadratic potential, for which a minimally coupled quintessence field can exhibit an oscillating equation of state {that remains in the quintessence regime}, $w\gtrsim-1$ \cite{Amin:2011hu,Bouhmadi-Lopez:2026wub,Jiang:2026cqh}. An oscillatory equation of state with phantom crossing can then be realized in modified-gravity frameworks \cite{Nojiri:2025low} or with quintom fields \cite{Feng:2004ff}. Recently, in Ref.~\cite{Nojiri:2025uew}, the condition for phantom crossing within EsGB gravity was realized by reconstructing the nonminimal coupling term as a function of the number of e-folds, $N=\log(a)$. Additionally, it has been shown that a non-dynamical field $\phi$ can behave as a non-relativistic dust particle when it couples to the Gauss-Bonnet function, which can also increase the effective mass of the field. Consequently, the energy density of the dark matter particle decays more slowly than $a^{-3}$, where $a$ is the scale factor. This scenario can exhibit acceleration in the low-redshift regime, similar to the CPL parametrization. However, unlike the reconstruction mechanism used to realize CPL-like behavior, we establish a framework based on a suitable choice of the potential and Gauss-Bonnet coupling function, for which the system exhibits oscillatory phantom crossing at the low redshift. We will demonstrate that, unlike the parametrization proposed in Ref.~\cite{Hussain:2026srf}, in which the non-transformed damping term can lead to a future singularity, no such future divergence arises in the physically motivated field-theoretic scenario. In other words, the oscillation occurs only over a finite epoch, and as the system transitions to the far future, the model approaches a stable de Sitter solution. 

We achieve this scenario by adopting a potential constructed from a combination of exponential and quadratic functions of the field $\phi$. Due to this combination, a minimum can be formed around which the slowly rolling scalar field oscillates. However, if the minimum of the potential is sufficiently shallow, the kinetic energy gained by the field while traversing the potential well can eventually drive the field away from the minimum. Thus, for a suitable value of the exponential parameter, the field exhibits oscillations only for a finite period {around the minimum, after which} the oscillations cease and the field approaches a de Sitter-like solution. In the absence of the Gauss-Bonnet coupling function, the oscillations do not cross the phantom divide; {the coupling, however, can} exert additional negative pressure, causing the equation of state to reach more negative values. Therefore, to demonstrate the effect of the nonminimal coupling only at late times, we consider a Gaussian-type functional form for the Gauss-Bonnet coupling function, which becomes nonzero only around a critical value of $\phi$ where the effective potential develops a minimum. We consider this particular form of the Gauss-Bonnet coupling function $f(\phi)$ {for the following reasons}. In Ref.~\cite{Hussain:2025vbo}, {model-independent dynamics of the system was realized by imposing the constraint $\ct=1$ on the tensor sound speed}. Under this constraint, the time derivative of the coupling function becomes proportional to the scale factor, $\dot{f}\propto a$. In such a scenario, it was shown that the model can produce an observationally viable matter- and radiation-dominated epoch only when the corresponding Gauss-Bonnet density at the present epoch is smaller than $10^{-20}$. {Similar constraints on the Gauss-Bonnet coupling function have also been obtained in} Refs.~\cite{Hussain:2024yee,Arora:2025ecj,Wang:2021kuw}. As the Gauss-Bonnet coupling function must remain very small to maintain $\ct=1$, a large value of the coupling can increase the Gauss-Bonnet density and eventually  adversely affect the early-Universe. Since the oscillatory phenomenon appears only at late times, we {choose} a Gaussian function whose contribution becomes significant only around the critical value of the field. This enables us to satisfy the tensor-speed constraint in the higher-redshift regime while investigating how it is affected by the oscillatory behavior of the field.

We study the effective behavior of the model at the background level using a dynamical-system stability framework and analyze the evolution of the cosmological parameters and dynamical variables. We obtain the critical points and apply linear stability theory to determine their stability conditions. For some critical points, linear stability theory becomes inconclusive; hence, in these cases, we {analyze their stability numerically as well as via} the center manifold theorem. As the dark-energy equation of state crosses the phantom divide multiple times, we linearly perturb the homogeneous background metric and {derive} the scalar and tensor propagation speeds, and determine the {no-ghost and gradient instability conditions}. We find that, for certain regions of the parameter space where the oscillation amplitude does not evolve toward increasingly negative values, the system does not exhibit ghost, and both sound speeds remain positive, indicating the absence of gradient instabilities and exponential growth of perturbation modes.

The rest of the paper is organized as follows. In Sec.~\ref{sec:background}, we derive the equations of motion and establish the Friedmann equation in the FLRW metric. We present a detailed analysis of the field behavior around the minimum of the potential, both in the absence and presence of the nonminimal coupling function, in Sec.~\ref{sec:theoretical_analysis_model}. In Sec.~\ref{sec:dynamics}, we study the stability and behavior of the model using the dynamical-system analysis framework and present a detailed discussion of the critical points and the reduced phase-space dynamics. In Sec.~\ref{sec:linear_perturbation}, we extend the analysis to linear perturbations of the metric, where we obtain the scalar and tensor sound speeds and determine the conditions under which the model realizes a phantom solution without ghost or gradient instabilities. Finally, we draw a brief conclusion in Sec.~\ref{sec:conclusion}.

\section{Background Equations}
\label{sec:background}
 The action governing the nonminimally coupled scalar field with the GB invariant $\EGB$ is given by
\begin{equation}
	S = \int d^4x\sqrt{-g}\left[
	\frac{R}{2\kappa^2}
	-\frac{1}{2}\nb_{\mu}\phi\nb^{\mu}\phi
	-V(\phi)
	-f(\phi)\,\EGB
	\right]
	+S_{\rm m}+S_{\rm r}\,,
	\label{eq:action}
\end{equation}
where $R$ is the Ricci scalar, $\kappa$ is the inverse Planck mass, $\phi$ is a quintessence scalar field with the potential $V(\phi)$ responsible for late-time cosmic acceleration, $S_{\rm m,r}$ is the action for the background pressureless matter and radiation components, which are minimally coupled to the gravitational sector. Here, $f(\phi)$ is the nonminimal coupling function, and $\EGB$ is 
\begin{equation}
	\label{eq:GB}
	\EGB \equiv R^2 - 4R_{\mu\nu}R^{\mu\nu}
	+ R_{\mu\nu\rho\sigma}R^{\mu\nu\rho\sigma}\,.
\end{equation}
In 4D, the scalar field coupling $f(\phi)\EGB$ contributes non-trivially and modifies the gravitational sector. Since the other constituents of the Universe are minimally coupled to gravity, the effect of the nonminimal coupling propagates to the other species gravitationally. The other species conventionally obey the standard fluid equations and are individually conserved, i.e., $\nb_{\mu}T^{\m}_{(m,r)}=0$, where $T^{\m}$ represents the energy-momentum tensor of the individual components. Assuming the Universe to be homogeneous, isotropic, and flat, the line element reads as
\begin{equation}
	ds^2 = -dt^2 + a(t)^2 d\vec{x}^2,\label{flrw_metric}
\end{equation}
where $a$ is the scale factor. Corresponding to this metric, the GB invariant becomes $\EGB = 24H^2 \left(\Hdot + H^2\right)$, where $H\equiv\dot{a}/a$ is the Hubble parameter, and an overdot denotes a derivative with respect to cosmic time. Varying the action with respect to the metric $g^{\m}$ yields
\begin{equation}
	G_{\m} = \kappa^2 \bigg(T_{\m}^{\phi}+ T_{\m}^{\rm m}+ T_{\m}^{\rm r} + T_{\m}^{\rm GB}\bigg),
\end{equation}
where the stress-energy tensors of different components are 
\begin{align}
	T_{\m}^{\phi} &= \nb_{\mu}\phi \nb_{\nu} \phi - \frac{1}{2}g_{\m} \bigg(\nb^{\alpha}\phi\nb_{\alpha}\phi + 2V(\phi)\bigg)\ ,\\	
	T_{\m}^{\rm m} &= \rho_{m} \ u_{\mu} u_{\nu} + P_{m}(u_{\mu} u_{\nu} + g_{\m}) \ ,\\
	T_{\m}^{\rm r} &= \rho_{r} \ u_{\mu} u_{\nu} + P_{r}(u_{\mu} u_{\nu} + g_{\m}) \ , \\
	T^{\rm GB}_{\m}& = -2\left(\nb_\mu \nb_\nu f\right)R+2g_{\mu\nu}\left(\nb_\rho \nb^\rho f\right)R  \nonumber\\&
	+4\left(\nb^\rho \nb_\nu f\right)R_{\mu\rho} 
	+4\left(\nb^\rho \nb_\mu f\right)R_{\nu\rho}-4\left(\nb^\rho \nb_\rho f\right)R_{\mu\nu} \nonumber\\
	&
	- 4 g_{\mu\nu}\left(\nb^\rho \nb^\sigma f\right)R_{\rho\sigma}+4\left(\nb^\rho \nb^\sigma f\right)R_{\mu\rho\nu\sigma}\ ,
\end{align}
and $u^{\mu}=(1,0,0,0)$ is the comoving four-velocity of the fluid, $\rho_{\rm m,r}$ represents the energy density of matter and radiation, and the corresponding pressure is denoted by $P_{\rm m,r}$. We assume matter to behave as a pressureless dust fluid, while the radiation pressure is $P_{\rm r} = \rho_{\rm r}/3$. In the flat FLRW metric given in Eq.~\eqref{flrw_metric}, the Friedmann equation becomes
\begin{eqnarray}
	3H^2 &=& \rDE + \rho_m + \rho_r  + 24H^3\dot{f}\ ,\label{frd_eq}
\end{eqnarray}
and the field equation becomes
	\begin{equation}
	\phidd + 3H\phidot + V_{,\phi} + f_{,\phi}\,\EGB = 0\,.
	\label{field_eqn}
\end{equation}
The equation of motion for other species is
	\begin{equation}
	\dot{\rho}_{\rm m,r} + 3 H (\rho_{\rm m,r} +P_{\rm m,r}) = 0\ . \label{matter_continuity}
\end{equation}
Corresponding to this metric, the field components, such as the energy density, pressure, and equation of state, become
\begin{equation}
	\label{eq:rhoDE}
	\rDE = \frac{1}{2}\phidot^2 + V(\phi) \,, P_{\phi} = \frac{1}{2}\phidot^2 - V(\phi), \ w_\phi = P_{\phi}/\rDE\ ,
\end{equation}
and GB density becomes
\begin{equation}
	\rho_{\rm GB} = 24 H^3 \dot{f}, \ \text{with}\ \  \EGB = 24H^2\!\left(\Hdot + H^2\right).
\end{equation}
Similar to the continuity equation obtained for the matter and radiation components in Eq.~\eqref{matter_continuity}, we can cast the field equation in a similar form as
\begin{equation}
	\dot{\rho}_{\phi} + 3 H(\rho_{\phi} + P_{\phi}) = -24H^2 f_{,\phi} \dot{\phi} (H^2 + \dot{H}^2)
\end{equation}
which allows us to define the effective pressure of the quintessence field due to the GB coupling as
\begin{equation}
	P_{\phi}^{\rm eff} = P_{\phi} + 8 H f_{,\phi} \dot{\phi} (H^2 + \dot{H}), \  w_{\phi}^{\rm eff} = \frac{P_{\phi}^{\rm eff}}{\rho_\phi}, \label{eos_eff_phi}
\end{equation}
and hence one can rewrite the above equation in a standard continuity relation as
\begin{equation}
	\dot{\rho}_{\phi} + 3 H (1+w_{\phi}^{\rm eff}) \rho_\phi = 0 \ .
\end{equation}
For the quintessence field, whose equation of state remains $w_{\phi}>-1$, the GB coupling can induce phantom behavior, allowing the effective equation of state to cross the phantom divide, $w\lesssim -1$, under suitable conditions. The time derivative of the Hubble parameter can be obtained by taking the time derivative of the Friedmann equation, Eq.~\eqref{frd_eq}, and using the field-fluid equation of motion, is given by
\begin{align}
	&\dot{H} = \frac{\mathcal{N}}{\mathcal{D}}, \text{where}, \
	\mathcal{N} = -\kappa^2 \Bigg[
	3\Bigg(
	32 H^3 \dot{\phi}\,f_{,\phi}
	+192 H^6 f_{,\phi}^{\,2} \nonumber\\ & +8H^2\Bigl(f_{,\phi}V_{,\phi}
	-\dot{\phi}^{\,2}f_{,\phi\phi}\Bigr)
	+\dot{\phi}^{\,2}
	\Bigg)+3\rmm+4\rr
	\Bigg], \nonumber
	\\
	\mathcal{D} &=
	6\left(
	1
	-8\kappa^2 H\dot{\phi}\,f_{,\phi}
	+96\kappa^2 H^4 f_{,\phi}^{\,2}
	\right). \label{Hdot}
\end{align}
With this one can determine the effective equation of state of the system and deceleration parameter as
\begin{equation}
	w_{\rm eff} = -1- \dfrac{2 \dot{H}}{3H^2},\ q= -1-\frac{\dot{H}}{H^2}\ .
\end{equation}

\section{Analytical Analysis of the Model}
\label{sec:theoretical_analysis_model}

We consider a potential consisting of an exponential function with a nearly flat region, allowing the field to initially undergo slow roll toward the minimum of the potential. This behavior is achieved by adding a quadratic function of $\phi^2$, such that the potential becomes
\begin{equation}
	V(\phi) = V_0 e^{-\kappa \lambda \phi} + \frac{1}{2}m_\phi^2 \phi^2\ . \label{newpotential}
\end{equation}
The potential is qualitatively similar to the potentials studied previously in Refs.~\cite{Amin:2011hu}, given by
\begin{equation}
	V(\phi) = \frac{m_{\phi}^2 M^2}{2} \bigg(\frac{(\phi/M)^2}{1+(\phi/M)^{2(1-\alpha)}}\bigg), \label{potential1}
\end{equation}
and the pseudo-Nambu-Goldstone-boson quintessence potential \cite{Hall:2005xb,Bouhmadi-Lopez:2026wub},
\begin{equation}
	V(\phi) = m_\phi^2 M^2 \bigg(1-\cos(\phi/M)\bigg), \label{potential2}
\end{equation}
where $M$ is a mass-dimensional constant and $m_\phi$ represents the mass of the quintessence field. These potentials exhibit a nearly flat regime in which the field rolls slowly before reaching the minimum of the potential, where the field begins to oscillate. Near the minimum, these potentials approximately exhibit a $\phi^2$ dependency. The behavior of the potential in Eq.~\eqref{newpotential} is shown in Fig.~\ref{fig:potential_variation} for different values of $m_\phi$ and $\lambda$. The minimum of the potential occurs at a small value of $\phi$ when the magnitude of the exponential contribution becomes comparable to that of the quadratic contribution, followed by a nearly flat region where the field initially rolls.
\begin{figure}[tbh]
	\resizebox{\columnwidth	}{!}{\includegraphics{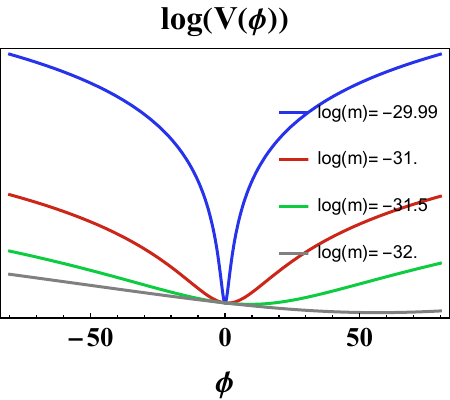}}
	\resizebox{\columnwidth	}{!}{\includegraphics{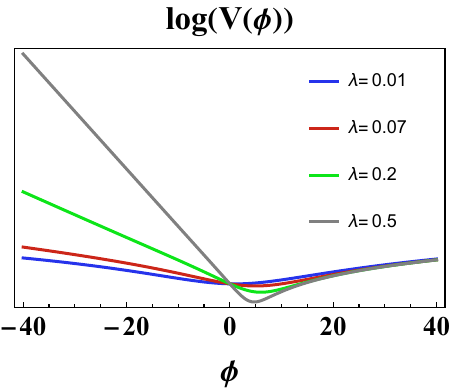}}	
	\caption{The scalar-field potential in Eq.~\eqref{newpotential} is shown on a logarithmic scale, with $V_0=10^{-60}$ and $\kappa=1$. In the upper panel, the scalar-field mass $(m_\phi=m)$ is varied logarithmically (base 10) while keeping $\lambda=0.01$ fixed. In the lower panel, the mass is fixed at $m_\phi=10^{-31}$, and the parameter $\lambda$ is varied.}
	\label{fig:potential_variation}
\end{figure}
Depending on the relative magnitudes of the coefficients of the exponential and quadratic terms, the location of the minimum can vary, with the minimum occurring away from $\phi\to 0$. In contrast, for the potential in Eq.~\eqref{potential1} the minimum always occurs around $\phi\to 0$, while periodic minima occur for the potential in Eq.~\eqref{potential2}. In addition to the coefficients of the potential, $(V_0,m_\phi)$, the exponential parameter $\lambda$ also plays an important role, as a smaller value of $\lambda$ decreases the depth of the minimum around $\phi\to 0$. Therefore, the combination of an exponential and a quadratic function exhibits behavior similar to that of the other potentials, while potentially leading to interesting dynamics because the depth of the potential can be regulated by the external parameter $\lambda$, unlike the other potentials, where different values of $M$ can directly rescale the value of $\phi$.

\subsection{Analysis without GB coupling $f(\phi)=0$}

At first, we study the oscillatory phase of the field in the absence of the GB coupling, i.e., $f(\phi)=0$. During the matter- or radiation-dominated phase, we consider the field energy density to be comparable to that of the cosmological constant, $\rho_{\phi}\sim\rho_{\Lambda}$, while the field remains in the quintessence regime with equation of state $w_{\phi}\sim -1$. Hence, the kinetic contribution remains suppressed relative to the potential energy, $\frac{1}{2}\dot{\phi}^2\ll V(\phi)$. Therefore, during these epochs, up to the low-redshift regime, the field $\phi$ does not undergo significant evolution from its initial value \cite{Amin:2011hu}. As the system enters the low-redshift regime, $z<1/3$, the matter density decays significantly and the Hubble parameter becomes sufficiently small compared with the characteristic energy scale of the potential, allowing the field $\phi$ to slowly roll toward the minimum of the potential, $V_{,\phi}=0$, satisfying
\begin{equation}
	\kappa \lambda V_0 = \frac{m_\phi^2 \phi_c}{e^{-\kappa \lambda \phi_c}},
\end{equation}
where $\phi_c$ represents the critical value of the field at which $V_{,\phi}=0$, provided that
\begin{equation}
	V_{,\phi\phi} = m_\phi^2 + \kappa \lambda m_\phi^2 \phi_c >0.
\end{equation}
Therefore, $(1+\lambda\kappa\phi_c)>0$, which implies $\kappa\phi_c<-1/\lambda$ for $\lambda<0$ and $\kappa\phi_c>-1/\lambda$ for $\lambda>0$. To examine the behavior of the field around the minimum of the potential at the homogeneous background level, we perturb the field to linear order around its minimum as
\begin{equation}
	\phi(t)=\phi_c+\delta\phi(t).
\end{equation}
Here, for the time being, we consider the field fluctuations to remain homogeneous, such that the fluctuations depend only on time. Expanding the field equation, Eq.~\eqref{field_eqn}, around the minimum, we obtain
\begin{equation}
	\ddot{\delta \phi} + 3 H \dot{\delta \phi} + \mph (\phi_c + \delta \phi) - \kappa \lambda V_0 e^{-\kappa \lambda \phi} = 0\ .
\end{equation}
Linearizing the exponential function, we get, 
\begin{align}
	\ddot{\delta \phi} + 3 H \dot{\delta \phi} + \mph (\phi_c + \delta \phi) - \mph \phi_c (1 - \kappa \lambda \delta \phi ) & = 0\ ,\\
	\ddot{\delta \phi} + 3 H \dot{\delta \phi} + \mph(1+\lambda \kappa \phi_c ) \delta \phi &= 0 .
\end{align}
The equation takes the form of a damped harmonic oscillator, where $3H$ acts as a time-dependent damping term. Thus, for an arbitrary value of $H$ around the minimum of the potential, we obtain the solution for $\delta\phi$ as
\begin{equation}
	\delta \phi = e^{-\frac{3 H t}{2}} \cos(\omega_\phi t + \varphi),
\end{equation} 
where $\varphi$ denotes a constant phase of the oscillation, and $\omega_\phi$ is the frequency of the oscillation, given by
\begin{equation}
	\omega_\phi = m_{\rm eff} \sqrt{1- \frac{9H^2}{4 m_{\rm eff}^2}}, \qquad m^{2}_{\rm eff} = \mph (1+\lambda \kappa \phi_c)\ .
\end{equation}
As long as $m_{\rm eff} > 3H/2$, the system oscillates, while the amplitude of the oscillation decays exponentially. The field's equation of state remains $w_{\phi} \gtrsim -1$, and no phantom crossing occurs, as also found in Refs.~\cite{Bouhmadi-Lopez:2026wub,Jiang:2026cqh}. Nevertheless, many standard and non-standard parameterized equations of state models, including CPL, Pade, Bell, and oscillatory models of the equation of state \cite{Hussain:2025nqy,Hussain:2026srf,Hussain:2026wss}, hint at the possibility of phantom crossing. Therefore, we study the current framework in the presence of the GB coupling to explore whether the oscillations can cross the phantom divide.

\subsection{Analysis with $f(\phi) \ne 0$}

To study the effect of oscillations and the phantom-crossing phenomenon, we consider a Gaussian form of the GB coupling centered around the critical field value $\phi_c$, where $V_{,\phi}=0$:
\begin{equation}
	\label{f_Gaussian}
	f(\phi) = \xi_0\,
	\exp\!\left[-\frac{\kappa^2 (\phi-\phi_c)^2}{2\sigma^2}\right],
\end{equation}
where $\sigma$ is the width parameter and $\xi_0$ is the coupling amplitude. The derivatives with respect to the field $\phi$ are given by
\begin{align}
	f_{,\phi}   &= -\frac{\kappa^2 (\phi-\phi_c)}{\sigma^2}\,f(\phi)\,,
	\label{eq:fphi}\\
	f_{,\phi\phi}&=  \frac{\kappa^4 (\phi-\phi_c)^2 - \kappa^2 \sigma^2}{\sigma^4}\,f(\phi)\,.
	\label{eq:fphiphi}
\end{align}
Therefore, as the system moves away from the minimum, the coupling weakens, and the system gradually approaches the minimally coupled regime. In order to examine the effect of the coupling, we consider the field equation as in the previous section. Due to the GB coupling, the effective mass of the field is further modified as
\begin{align}
	\ddot{\delta \phi} + 3 H \dot{\delta \phi} + \mph(1+\lambda \kappa \phi_c ) \delta \phi - 24\frac{\kappa^2 f}{ \sigma^2} H^4 \left(\frac{\dot{H}}{H^2}+1\right) \delta \phi &= 0 
\end{align}
and the frequency becomes
\begin{equation}
	\omega_\phi = m_{\rm eff} \sqrt{1-\frac{9H^2}{4m_{\rm eff}^2}}, \qquad
	m_{\rm eff}^2 = \mph \left(1+\lambda \kappa \phi_c-\gamma H^2\right),
\end{equation}
where
\begin{equation}
	\gamma = 24\frac{\kappa^2 f H^2}{\mph \sigma^2}
	\left(\frac{\dot{H}}{H^2}+1\right),
\end{equation}
and the field fluctuation satisfies
\begin{equation}
	\delta\phi(t) = e^{-\frac{3Ht}{2}}\cos(\omega_\phi t+\varphi).
\end{equation}
In this case, the system oscillates as long as
\begin{equation}
	\gamma H^2 < 1+\lambda\kappa\phi_c,
\end{equation}
provided that $m_{\rm eff}>3H/2$. Corresponding to this oscillatory scenario, the effective equation of state of the field in Eq.~\eqref{eos_eff_phi} becomes
\begin{equation}
	w_{\phi}^{\rm eff} = \frac{\frac{1}{2} \dot{\phi}^2 - V(\phi) + 8 H f_{,\phi} \dot{\phi} (\dot{H} + H^2)}{\frac{1}{2} \dot{\phi}^2 + V(\phi)}\ . \label{eos_eff_phi_val}
\end{equation}
During the transition epoch between DM and DE, particularly in the very low-redshift regime where the field starts to oscillate, the potential energy of the field dominates over its kinetic energy. Therefore, from the above expression, we can neglect the $\dot{\phi}^2$ term, and the equation of state (EoS) becomes
\begin{equation}
	w_{\phi}^{\rm eff} \approx -1 - \dfrac{8H}{V} \frac{\kappa^2 \delta \phi}{\sigma^2} f \dot{\delta \phi} H^2 \left(\frac{\dot{H}}{H^2} + 1\right)\ .
\end{equation}
The deceleration parameter of the system is defined as $q = - \left(\frac{\dot{H}}{H^2} + 1\right)$. Thus, the equation of state becomes
\begin{equation}
	w_{\phi}^{\rm eff} \approx -1 + \dfrac{8H}{V} \frac{\kappa^2 \delta \phi}{\sigma^2} f \dot{\delta \phi} H^2 q \ .
\end{equation}
For $\Lambda$CDM, the transition regime is characterized by $q<0$. Since $|\delta\phi|$ and $|\dot{ \delta \phi}|$ exhibits oscillatory behavior while the remaining parameters in the above expression are positive, the effective equation of state of the scalar field can cross the phantom divide for $f\ne0$ with $q<0$.

%However, in Ref.~\cite{Hussain:2025vbo}, where $f$ takes a particular functional form imposed by the constraint $\ct=1$, the overall magnitude of the GB coupling becomes extremely small, and hence the system does not cross the phantom divide.

\section{Dynamical System and Stability Analysis}
\label{sec:dynamics}
% ============================================================

%TODO: Now lets change from here. 

In this section, we analyze the stability of the model at the background level using the dynamical-systems stability framework. We do this by transforming the field equation Eq.~\eqref{field_eqn} and the Friedmann equation Eq.~\eqref{frd_eq} into a set of coupled first-order differential equations, also known as an autonomous system, by defining dimensionless dynamical variables. The dimensionless variables are
\begin{multline}
	x    = \frac{\kappa \phidot}{\sqrt{6}\,H}\,,
	y    = \frac{\kappa^2 V_0 e^{-\kappa \lambda \phi}}{3\,H^2}\,,
	A  = \frac{\kappa^2 m^2 \phi^2}{6 H^2}, z= \kappa \phi \\
	u   =  8 \kappa^2 H^2 f\,,
	\Or  = \frac{\kappa^2\rho_{\rm r}}{3H^2}\,, \Om = \frac{\kappa^2 \rho_{\rm m}}{3H^2}, z_c = \kappa \phi_c\ .
	\label{dyn_variable}
\end{multline}
Here, we normalize the field velocity $\dot{\phi}$ with respect to the Hubble parameter and capture the corresponding dynamics through the variable $x$, while $y$ is defined by normalizing the exponential part of the potential with respect to $H$. The quadratic part of the potential, normalized with respect to $H$, is captured through $A$. In order to close the system and obtain the corresponding critical points, we define an additional variable $z$ that captures the field $\phi$. The dynamics of the GB coupling parameter is captured by $u$, which is proportional to $f$ rather than $\dot f$, as it appears in Eq.~\eqref{frd_eq}. Since, in the current study, we choose a Gaussian function for $f$, defining the dynamical variable corresponding to the original function $f$ sufficiently captures its essence and closes the system without requiring additional variables. The fractional energy-density parameters $\Omega_{\rm r}$ and $\Omega_{\rm m}$ describe the evolution of the radiation and matter densities, respectively. The variable $z_c$ represents the constant value of $\phi$ around which the GB coupling function Eq.~\eqref{f_Gaussian} takes a non-zero value. In terms of these variables, the Friedmann equation, which provides a constraint on the dynamical system, becomes
\begin{align}
	1 &= x^2 + y+A + \Or + \Om - \frac{\sqrt{6} u x (z-z_c)}{\sigma^2}\ .
	\label{eq:constraint}
\end{align}
Using this equation, the dynamics of one of the variables, namely the radiation density, can be expressed in terms of the other dynamical variables as $\Or=1-x^2-y-A-\Om-\frac{\sqrt{6} u x (z-z_c)}{\sigma^2}$. Since the radiation density must always be positive and bounded within $0\le \Or \le 1$, any solution that violates this condition cannot be considered physically viable. Therefore, the Friedmann equation not only provides a physical constraint on the system but also reduces the dimensionality of the independent variables required to close the system. The other relevant parameters, including the field density, GB density, and effective equation of state, can be expressed in terms of the dynamical variables as
\begin{multline}
	\Ode  = x^2 +y +A, \  \  \Og =  - \frac{\sqrt{6} u x (z-z_c)}{\sigma^2}, \\
	\weff \equiv -1 - \frac{2\Hdot}{3H^2}\  ,
	\label{density_def}
\end{multline}
where $\frac{\dot{H}}{H^2} \equiv \frac{\mathcal{N}}{\mathcal{D}}$, with
\begin{multline}
	\mathcal{N} = \sigma ^2 u \bigg(6 A \left(z-z_c\right)+z \bigg(4 \sqrt{6} x \left(z-z_c\right)+3
	\lambda  y \left(z_c-z\right) \\-6 x^2\bigg)\bigg)-3 u z \left(z-z_c\right)^2
	\left(u-2 x^2\right)+\sigma ^4 (-z) \left(3 \Omega _m+4 \Omega_r+6 x^2\right)\ , \\
	\mathcal{D} = z \left(3 u^2 \left(z-z_c\right)^2+2 \sqrt{6} \sigma ^2 u x \left(z-z_c\right)+2
	\sigma ^4\right)\ . \label{dotH}
\end{multline}
Similarly, the effective equation of the state Eq.~\eqref{eos_eff_phi_val} of the field becomes
\begin{equation}
	w_{\phi}^{\rm eff}  = \dfrac{x^2 -A-y  -\frac{\sqrt{\frac{2}{3}} u x (z-\text{zc}) \left(-\epsilon +1\right)}{\sigma ^2 }}{x^2+A+y} \ ,\label{eos_phi_eff_dyn}
\end{equation}
where $\epsilon\equiv-\Hdot/H^2$. The autonomous equations of the system are
\begin{eqnarray}
	x' & = &\frac{\sqrt{\frac{3}{2}} u z \overset{.}{H}}{H^2 \sigma
		^2}-\frac{\sqrt{\frac{3}{2}} u z_c \overset{.}{H}}{H^2 \sigma ^2}-\frac{x
		\overset{.}{H}}{H^2}-\frac{\sqrt{6} A}{z} +\frac{\sqrt{\frac{3}{2}} u z}{\sigma
		^2}-\nonumber \\
	&&\frac{\sqrt{\frac{3}{2}} u z_c}{\sigma ^2}  -3 x+\sqrt{\frac{3}{2}}
	\lambda  y \label{x_prime},\\
	y' & = & \left(-\sqrt{6}\right) \lambda  x y-\frac{2 y \dot H}{H^2},\\
	A' & =& \frac{2 \sqrt{6} A x}{z}-\frac{2 A \dot H}{H^2} \label{A_prime} , \\
	z' & =& \sqrt{6} x, \label{z_prime}\\
	u' &=& \frac{2 u \overset{.}{H}}{H^2}-\frac{\sqrt{6} u x z}{\sigma ^2}+\frac{\sqrt{6} u x
		z_c}{\sigma ^2},  \label{u_prime}\\
	\Om' & =& -\frac{2 \Om \overset{.}{H}}{H^2}-3 \Om\ . \label{om_prime}
\end{eqnarray}
Here, the prime denotes the derivative with respect to the $e$-folding parameter, $dN = H dt$. Six variables are required to close the system; therefore, the dimensionality of the phase space is six. In the autonomous system, we see that the variable $z$ appears in the denominator of $x'$ and $A'$, and hence the equations diverge as $z \to 0$. However, the system may remain non-singular at this point if the potential parameter $A \to 0$. To analyze the stability of the model, we obtain the critical points corresponding to the autonomous system of equations. The critical points and their corresponding stability conditions are summarized in Tab.~\ref{tab:critical_points}.

\subsection{Critical Points and Stability Analysis}
\begin{table}[t]
	\resizebox{\columnwidth}{!}{
	\begin{tabular}{l c c c c c r}
		\hline\hline
		Point & $x_{\star}$ & $y_{\star}$ & $A_{\star}$ & $z_{\star}$ & $u_{\star}$ & $\Omega_{\rm m\star}$ \\
		\toprule
		
		$P_1$ & $0$ & $\frac{2 \sigma ^2-u z^2+u z \  z_c}{\sigma ^2 (\lambda  z+2)}$ & $\frac{z \left(\lambda  \sigma ^2+u z-u \ z_c\right)}{\sigma ^2 (\lambda  z+2)}$ & Any & Any &$0$ \\
		$P_2$ & $0$ & $1$ & $0$ & Any & $-\frac{\lambda  \sigma ^2}{z-z_c}$ & $0$ \\
		$P_{3}$ & 0 & 0 & 0 &  Any & 0 & 0\\
		
		$P_{4}$ & 0 & 0 & 0 &  Any & 0 & 1\\
		\bottomrule		
	\end{tabular}}
\resizebox{\columnwidth}{!}{
\begin{tabular}{l | c |c |r}
	\toprule
	Point & Eigenvalues & Stability conditions &  $\Omega_{\phi}$ \\
\toprule
	$P_1$ & Fig. \ref{fig:point12_stability} & hyperbolic & 1 \\
	\bottomrule
	$P_2$ &
	Fig. \ref{fig:point12_stability} & non-hyperbolic & 1 \\
	
	\bottomrule
	$P_{3}$ & $\{0,-4,-1,1,4,4\}$ & \makecell{non-hyperbolic \&\\ Unstable}  & 0\\
	
	\bottomrule
	$P_{4}$ & $\left\{0,-3,-\frac{3}{2},-1,3,3\right\}$ & \makecell{non-hyperbolic \&\\ Unstable}  & 0\\
	\bottomrule
	\bottomrule
	\end{tabular}}
	\caption{The coordinates of the critical points of the system are labeled by $\star$. The point $P_{1}$ can take arbitrary values of $(A,z, u)$. The eigenvalues are computed at the corresponding critical points.}
	\label{tab:critical_points}
\end{table}

\begin{figure}
	\resizebox{\columnwidth}{!}{\includegraphics{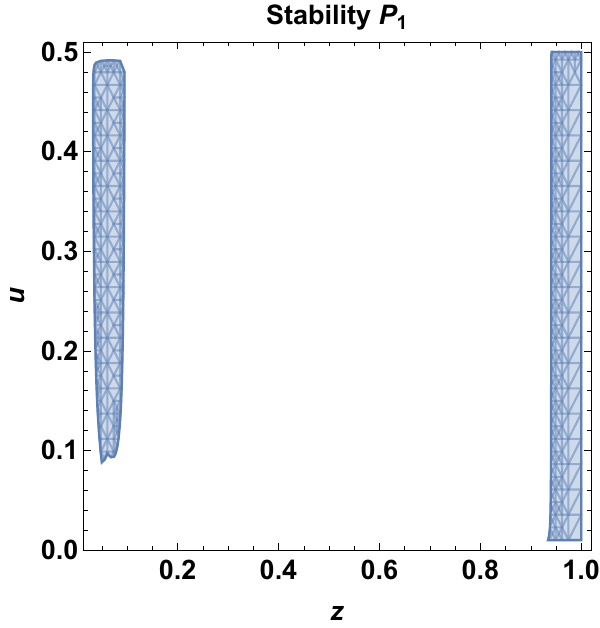}\includegraphics{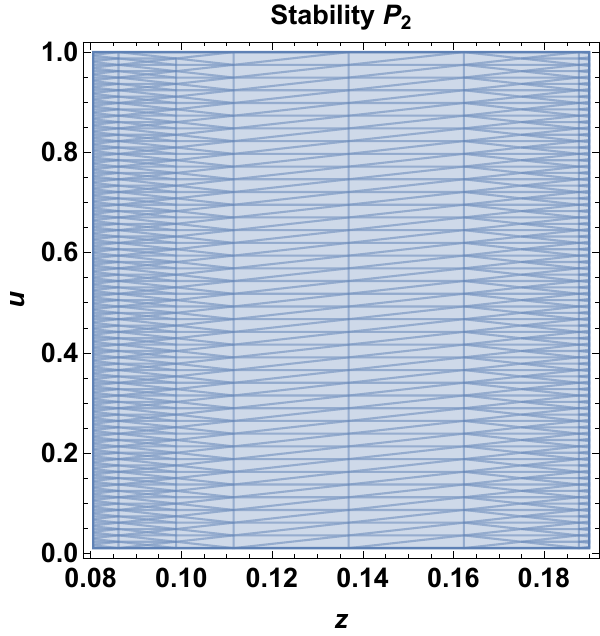}}
	\resizebox{\columnwidth}{!}{\includegraphics{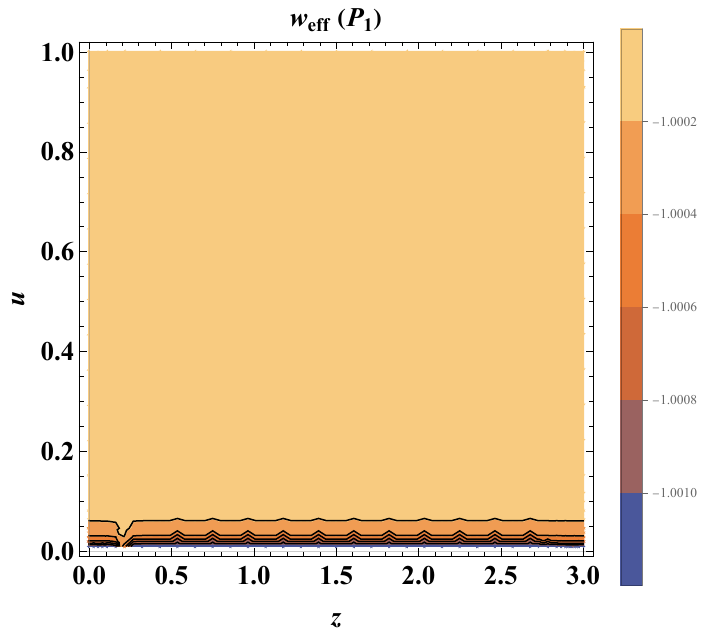}\includegraphics{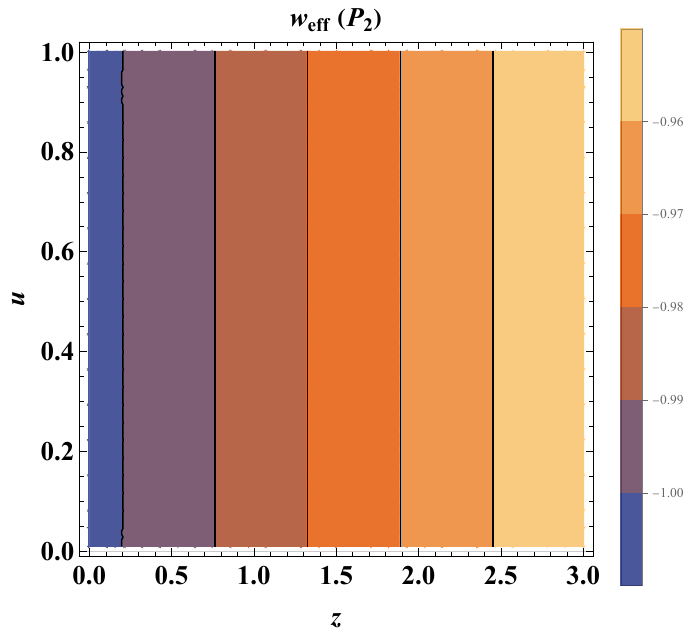}}
	\caption{The stability of the critical points for $\lambda = 0.1$, $\sigma = 0.01$, and $z_c = 0.2$. The point $P_{1}$ is a hyperbolic point, for which all eigenvalues have non-zero real parts, whereas $P_{2}$ is a non-hyperbolic point, for which one of the eigenvalues has a vanishing real part. Thus, in the parameter space shown above, $P_{2}$ has eigenvalues with negative real parts and one eigenvalue with a vanishing real part. Consequently, linear stability analysis is insufficient to determine the characteristics of this point. The effective equation of state at $P_{1}$ does not vary significantly with $\lambda$, $\sigma$, and $z_c$, whereas for $P_{2}$, $w_{\rm eff}$ increases with increasing $\lambda$ and $z_c$. For different values of the model parameters, the overall contour remains similar.}
	\label{fig:point12_stability}
\end{figure}

The autonomous system of equations, Eqs.~\eqref{x_prime}--\eqref{om_prime}, produces four critical points. In this autonomous system, Eq.~\eqref{z_prime} vanishes only for $x=0$, which causes $z$ to remain constant at the critical points. Therefore, the remaining critical points are calculated by fixing $x=0$, as reported in Tab.~\ref{tab:critical_points}. We label these critical points by $P_i$ and obtain their stability conditions by linearizing the right-hand side of the autonomous equations around each critical point using a Taylor-series expansion. From this expansion, we construct a $6\times6$ Jacobian matrix and evaluate its eigenvalues, which are tabulated in Tab.~\ref{tab:critical_points}.

If the real parts of all eigenvalues are non-zero, the point is hyperbolic, and its asymptotic behavior can be determined from the signs of the real parts. If all real parts are negative (positive), the point is asymptotically stable (unstable), whereas if they have mixed signs, the corresponding point is a saddle point. A saddle point is characterized by trajectories that approach the point along some directions in phase space and depart from it along others. If all trajectories in the relevant phase space approach and asymptotically settle at a point, the point is stable, whereas if trajectories repel from the point, it is unstable \cite{wainwright1997dynamical,Bahamonde:2017ize,Roy:2022fif}. The linear stability analysis fails when one or more of the real parts of the eigenvalues vanish while the remaining eigenvalues have negative real parts. To determine the characteristics of such points, we either rely on numerical evolution or employ more mathematically advanced techniques, such as the Center Manifold Theorem (CMT) \cite{B_hmer_2012}.

Finally, the physical significance of the critical points can be determined from the corresponding values of the effective equation of state, $w_{\rm eff}$, and the fractional energy densities at each point. A point is classified as accelerating (quintessential) if the corresponding $\eos$ satisfies $-1\le\eos\le-1/3$, phantom if $\eos<-1$, matter dominated when $\eos\sim0$, and radiation dominated when $\eos\sim1/3$.

{$\bullet$ \bf \boldmath Point $P_1$:}  The critical point is scalar-field dominated, with $\Ode = 1$. At this point, both potential variables take finite values whose sum satisfies $(y_{\star} + A_{\star} = 1)$. The non-minimal coupling variable $u_{\star}$ can take any value at this point. The eigenvalues evaluated at this critical point depend on the free parameters and can take different values for $(z,u)$. We plot the real parts of the eigenvalues in the region where they are negative for the benchmark values $\lambda = 0.1$, $\sigma = 0.01$, and $z_c=0.2$, and display this region in the $(u,z)$ parameter space. The point remains stable in the shaded regions, which occur around $z\sim 0.1$ and $z\sim 1.0$. In the white region, one of the eigenvalues has a vanishing real part. Therefore, the point can be either hyperbolic or non-hyperbolic depending on the asymptotic value of $z$. When the point becomes non-hyperbolic, its stability cannot be determined using the standard linear stability theorem. We therefore infer the stability in these regions by numerically solving the autonomous system in the vicinity of the corresponding critical points. Besides the numerical analysis, we also outline a general theory of CMT in Appendix~\ref{app:cmt}.

{$\bullet$ \bf \boldmath {Point $P_2$:}} The critical point is also scalar-field dominated; however, the potential $A_{\star}$ vanishes, and the variable corresponding to the exponential potential dominates. The variable $z$ can take any value, while $u_{\star}$ takes a specific value that depends on the model parameters $(\lambda, \sigma,z_c)$. The point becomes singular at $z_c$; however, the constraint equation Eq.~\eqref{eq:constraint} remains finite, since $\Omega_{\rm GB} \propto x$. Thus, at the asymptotic point, the contribution from the quadratic part of the potential vanishes. Consequently, the field does not oscillate asymptotically and saturates to the de-Sitter phase with $\eos = -1$. The point is non-hyperbolic, as one of the real parts of the eigenvalues is zero, and we illustrate the real parts with $\lesssim 0$ condition in Fig.~\ref{fig:point12_stability}. We therefore infer the stability of the model using numerical evolution. At this point, $\eos \gtrsim -1$ and increases as $u>0.04$. Thus, large values of $u$, irrespective of $z$, tend to drive the equation of state close to de-Sitter solution. 

%However, in this range, the real part of the eigenvalues becomes positive, and the point becomes unstable. Therefore, the system cannot asymptotically settle at higher values of $z$.

{$\bullet$ \bfseries \boldmath Point $P_{3}$:} At this point, all coordinates vanish except $z$, and from the constraint equation \eqref{eq:constraint}, the radiation fractional density $\Or = 1$ dominates, with the corresponding $\eos = 1/3$. At this point, only the variable $z$ can take arbitrary values; however, the regime remains unaffected since the kinetic component vanishes. One of the eigenvalues is zero, while the remaining eigenvalues have real parts with alternating signs. Since the eigenvalues include a positive real part, the point is always a saddle point, and it is therefore not necessary to go beyond linear stability analysis to determine its stability.

{$\bullet$ \bfseries \boldmath Point $P_{4}$:} At this point, $\Om = 1$, while the remaining variables vanish except $z$. The value of $z$ does not affect the matter-dominated regime, and the corresponding $\eos = 0$. Therefore, this point represents the matter-dominated phase. The point also remains a saddle point, as the non-vanishing real parts of the eigenvalues have alternating signs, while one of the eigenvalues vanishes.

\subsection{Reduced Phase space and Numerical simulation}

\begin{figure*}
	\begin{tabular}{lll}
		(a.)  & (b.) & (c.)\\
		{\includegraphics[scale=0.4]{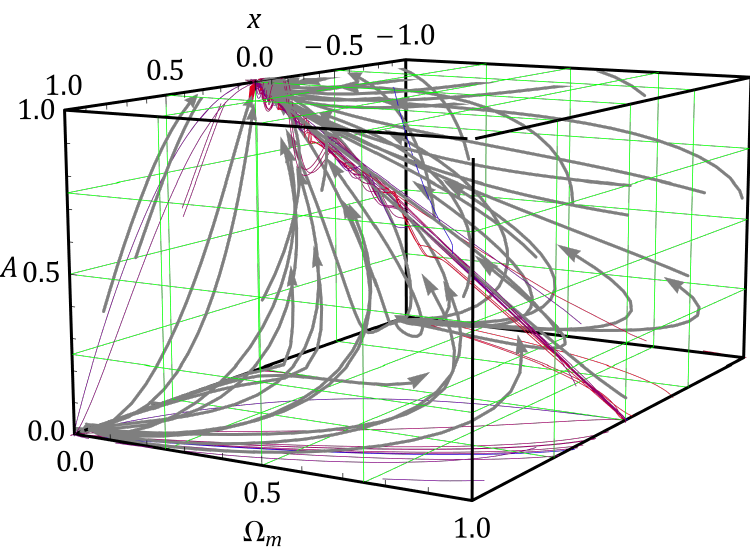}} &  {\includegraphics[scale=0.4]{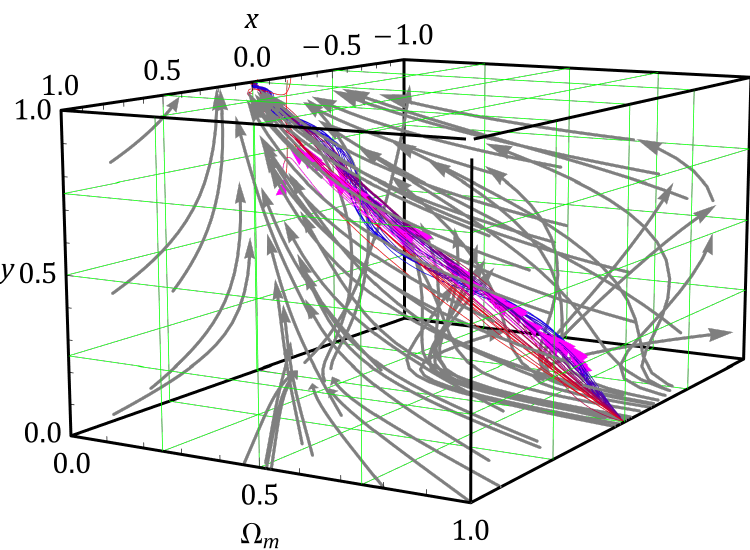}} & {\includegraphics[scale=0.4]{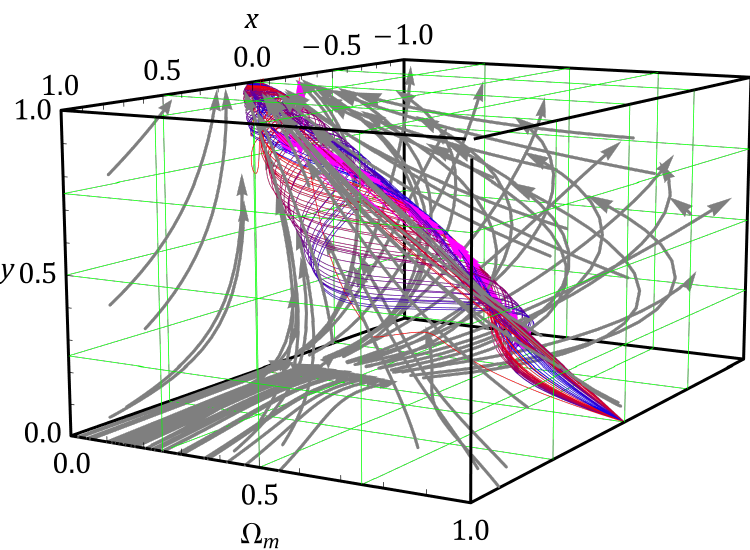}}
	\end{tabular}
	
	\caption{The reduced 3D phase space of the system in the coordinates $(x,\Om,A)$ and $(x,\Om,y)$, obtained by solving the autonomous system of equations, Eqs.~\eqref{x_prime}--\eqref{om_prime}, and shown by the blue and red curves. We set the initial conditions deep in the radiation-dominated regime and vary both the initial conditions and model parameters over the specified ranges. The gray curves are obtained by varying the algebraic expression of the corresponding variables while fixing the other variables to specified values. (a) The gray curves are obtained by fixing $y = 10^{-4}, u = 10^{-3}, z = z_c = 5, \sigma = 0.2, \lambda=0.01$. For the other curves, we vary $\log_{10} y_0 = [-50,-40], \log_{10}A_0 = [-33,-29], z_0 = [-6,6], u_0 = [-10^{-3},0.1], \lambda = [0.01,0.5]$, $\sigma \in [0.09,0.1]$, and $z_c = [-5.0,5.0]$, with fixed $x_0 = 10^{-30}$. We show only those solutions that satisfy $0<\Ode<1$ at $N=0$. (b) The gray curve is obtained by fixing $A=0.2, u=0.1, z=1.3,z_c=1.2, \sigma=0.099, \lambda=0.01$, while the other curves are obtained by varying $x_0=10^{-30}, \log_{10}y_0=[-34,-30.0], \log_{10}A_0=[-31,-29.3], z_0=[1.0,1.4], u_0=[0.1,0.2], \lambda=[0.01,0.3], \sigma_0=[0.09,0.1], z_c=[1.0,1.2]$. (c) The gray curve corresponds to the fixed values $A=0.002, u=0.05, z=1.3,z_c=1.2, \sigma=0.099, \lambda=0.01$, while the other curves are obtained by varying $x_0=10^{-30}, \log_{10}y_0=[-31.5,-29.9], \log_{10}A_0=[-29.9,-29.6], z_0=[1.0,1.4], u_0=[0.15,0.2], \lambda=[10^{-3},0.3], \sigma_0=[0.098,0.099], z_c=[1.0,1.2]$. The corresponding variations of each variable and cosmological parameters for (a), (b), and (c) are shown in Fig.~\ref{fig:numerical_simulation_phase_space}.}
	
	\label{fig:phase_space}
\end{figure*}
The system exhibits a 6-dimensional phase space; therefore, tracing the trajectories in the full phase space is not feasible. We therefore trace the phase-space trajectories in the lower-dimensional sub-spaces corresponding to the field and fluid variables, $(x,A,\Om)$ and $(x,y,\Om)$, as shown in Fig.~\ref{fig:phase_space}. The trajectories are obtained by solving the autonomous system of equations with initial conditions set deep in the radiation-dominated regime. The ranges of the initial conditions are given in the caption of the figure. Additionally, we plot the algebraic expressions corresponding to the right-hand sides of equations $(x',A',\Om')$ and $(x',y',\Om')$ by fixing the remaining variables and model parameters, shown in gray.

The vector trajectories obtained from the algebraic expressions span a broad region of the phase space, whereas the solutions of the differential equations trace only those trajectories that produce a matter-dominated phase followed by late-time dynamics dominated by the scalar field and remain non-divergent in the asymptotic past or future. We also show the evolution of each dynamical variable together with the cosmological parameters $(\Ode,\Om,\eos,H)$ in Fig.~\ref{fig:numerical_simulation_phase_space}. This individual evolution helps us decode the behavior of the model and provides additional insight into the phase-space representation.

In Fig.~\ref{fig:phase_space}a, we show the trajectory behavior in $(x,A,\Om)$. Here, we choose the initial condition, $\log_{10} y_0 = [-50,-40]$, which is much smaller than the $\log_{10}A_0 = [-33, -29]$, consequently making the exponential contribution much smaller than the quadratic contribution $A_0$ throughout the evolution, as shown in Fig.~\ref{fig:numerical_simulation_phase_space}a. In this case, we consider a broad range of initial conditions and apply a filter such that only solutions that produce an accelerating universe with a positive field energy density are retained. We consider $0<\lambda<0.5$, together with a relatively narrow range of $u_0$. Larger values of $u_0$ lead to stiffness in the numerical solver, as the dynamical variables become divergent.

For this range of parameters, when the exponential contribution is very small, the quadratic contribution plays the dominant role. As the system evolves toward lower redshifts, the field undergoes rapid oscillations, during which $\eos$ crosses $-1$ and can reach substantially higher negative values. In this case, $z$ does not exhibit significant deviation, while $u$ becomes finite only near $N=0$. Thus, the GB coupling enhances the oscillation amplitude of $\eos$, allowing it to reach more negative values and cross the phantom divide multiple times. As the system evolves toward the future, the solution asymptotically approaches $\eos=-1$. A similar late-time oscillatory behavior with excursions into the phantom regime has also been observed in the dark-energy parametrization studied in Ref.~\cite{Hussain:2026srf}.

In the phase space, some of the red and blue trajectories originate near $x\sim0$ or $x\sim1$, move toward $\Om\sim 1$, corresponding to the matter-dominated critical point $P_{4}$, and are eventually attracted toward $A\sim1$, corresponding to $P_{1}$. We do not label these points directly in the reduced phase space because it does not represent the full six-dimensional system. Moreover, their stability can be inferred numerically by observing the behavior of each dynamical variable as shown in Fig.~\ref{fig:numerical_simulation_phase_space}a. We find that, for a broad range of initial conditions, the variables asymptotically approach $P_{1}$, where $A$ dominates while the other dynamical variables $(y,u,z)$ remain non-zero and asymptotically saturate to a finite value without exhibiting any divergence. This behavior indicates that $P_{1}$ acts as a stable attractor for this range of parameters. We have also traced the tensor sound speed $(\ct)$ and adiabatic scalar sound speed $(\cs)$, which we will discuss in the next section.

We plot the phase-space trajectories with $(x,y,\Om)$ variables for different initial conditions in panels Fig. (\ref{fig:phase_space}b) and (\ref{fig:phase_space}c), with the corresponding numerical evolution of the parameters shown in Fig.~\ref{fig:numerical_simulation_phase_space}(b,c), respectively. In particular, we consider comparable ranges of $\log_{10}y_0=[-34,-30]$ and $\log_{10}A_0=[-31,-29.3]$. In these cases, both $y$ and $A$ attain finite values at $N=0$; however, $y>A$, and $\eos$ exhibits oscillatory behavior. Most of the initial conditions lead to solutions in which the $\eos$ oscillations remain above $-1$. This similar non-phantom oscillatory behavior has been reported in Ref.~\cite{Jiang:2026cqh} for the minimally coupled quintessence field with potential Eq.~\eqref{potential1}. However, some initial conditions for which $A$ takes a somewhat large positive value with larger $u$, exhibit larger negative values of $\eos$. In these scenarios, the variable $y$ dominates over $A$, and the oscillation disappears in the late-time epoch, leading to a stable de Sitter solution. This case highlights that both the critical points $P_{1,2}$ can asymptotically become stable points for different initial conditions; however, the physical description of the system remains the same. Upon examining the behavior of the trajectories in blue and red, we find that most of the trajectories corresponding to Fig.~\ref{fig:phase_space}(b,c) originate near $(x,y,\Om)=(0,0,0)$, evolve toward $\Om\sim1$, rendering a saddle matter-dominated epoch, and eventually stabilize at $(0,1,0)$, the field-dominated point where the exponential potential dominates. Thus, at the background level, the model yields a stable de-Sitter attractor while exhibiting oscillatory behavior during the transient DM--DE transition epoch around $N=0$. 

For each case, we plot the Hubble function on a logarithmic scale and compare its evolution with $\Lambda$CDM, showing that the deviation emerges only at low redshift, while both models remain consistent at higher redshift.

	\begin{figure*}
		\begin{tabular}{l}
			(a.)\\
			\resizebox{0.8\linewidth}{0.3\textheight}{\includegraphics{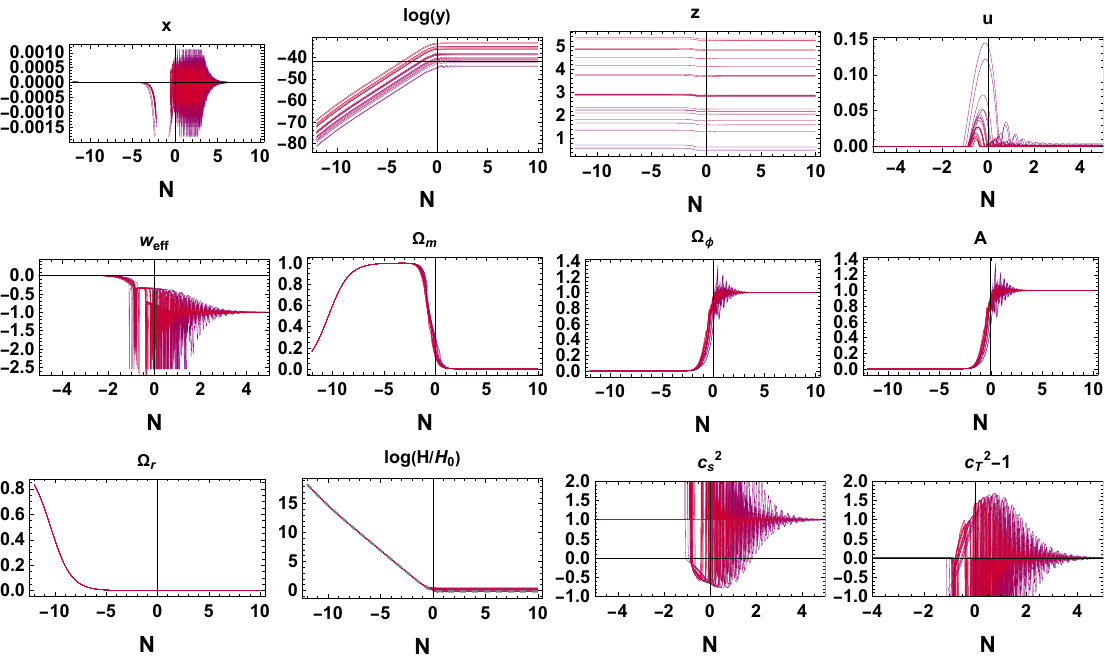}}\\
			(b.)\\
			\resizebox{0.8\linewidth}{0.3\textheight}{\includegraphics{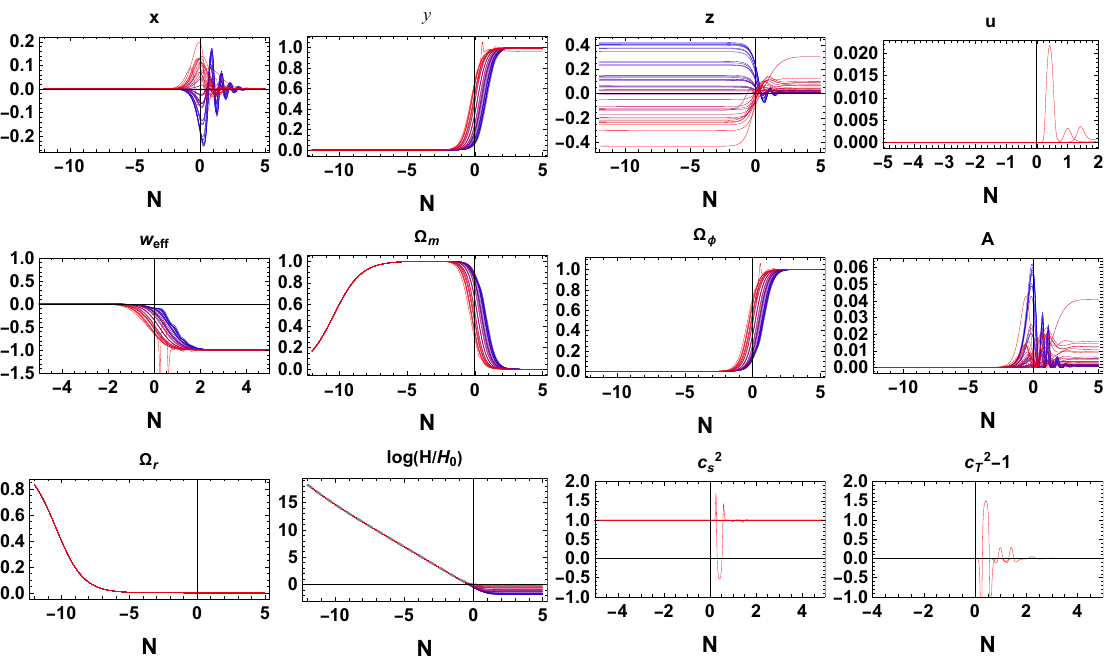}}\\
			(c.)\\
			\resizebox{0.8\linewidth}{0.3\textheight}{\includegraphics{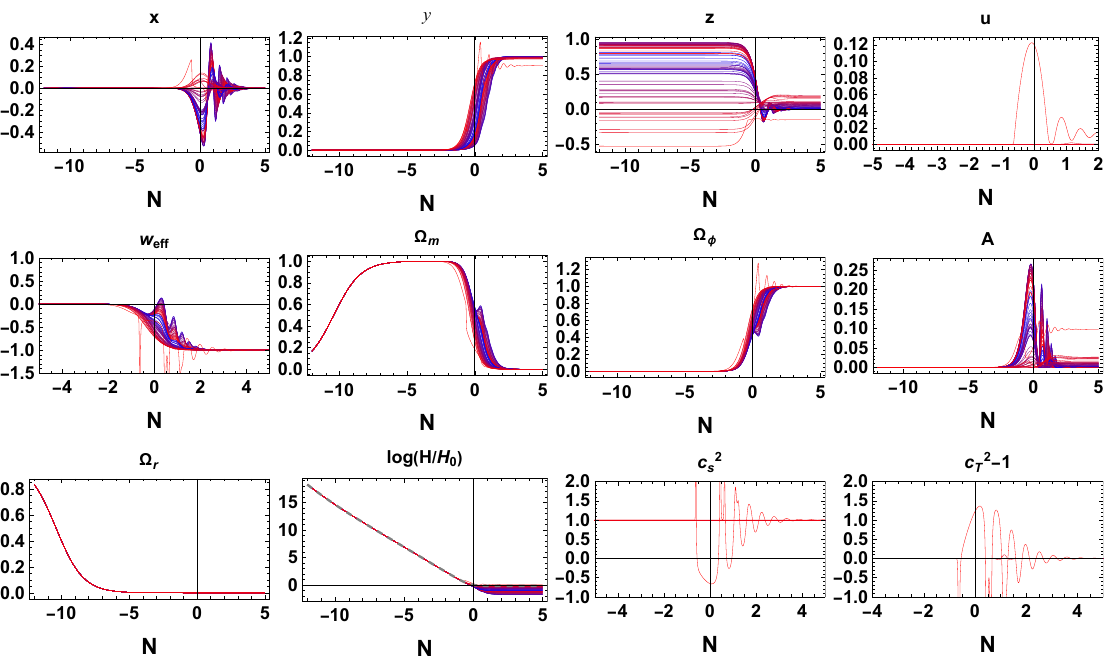}}
			
		\end{tabular}
	
	\caption{The numerical evolution of the cosmological parameters for initial conditions set deep in the radiation-dominated regime at $N\sim-20$. The initial conditions and behavior of the trajectories in the reduced phase space are discussed in Fig.~\ref{fig:phase_space}. The Hubble parameter is shown on a logarithmic scale as $H/H_0$, with the gray dashed line corresponding to the $\Lambda$CDM model.}
	\label{fig:numerical_simulation_phase_space}
\end{figure*}

\section{Linear Perturbation and Model's stability}
\label{sec:linear_perturbation}

In the previous section, we determined the stability of the model at the background level. However, the stability condition does not guarantee the absence of ghost or gradient instabilities at the level of linear perturbations. To obtain the corresponding conditions, we consider scalar and tensor perturbations of the background FLRW metric at first order. The perturbed line element can then be written as \cite{Bassett:2005xm,Tsujikawa:2006ph}
\begin{multline}
	ds^2=-(1+2 A) d t^2+2 a \partial_i B d x^i d t+a^2\bigg[(1+2 \psi) \delta_{i j}\\ +2 \pp_{ij} E+2 h_{i j}\bigg] d x^i d x^j,
\end{multline}
where $A$, $B$, $\psi$, and $E$ are scalar perturbation quantities, while $h_{ij}$ represents the tensor perturbation\footnote{It should be noted that $A$ does not denote the dynamical variable defined in Eq.~\eqref{dyn_variable}.}. Corresponding to this perturbed metric, the tensor sound speed $\ct$ and scalar sound speed $\cs$ can be expressed as \cite{Tsujikawa:2006ph} assuming $\kappa^2 =1$
\begin{eqnarray}
	\ct &=& \frac{1-8\Hdot f_{,\phi}\phidot - 8H^2f_{,\phi\phi}\phidot^2}
	{1-8\Hdot f_{,\phi}\phidot}\,,\\
	\cs&=&\bigg((1-8 H \dot{f})\left[(1-8 H \dot{f}) \dot{\phi}^2+96\left(H^2 \dot{f}\right)^2+128 H^2 \dot{H} \dot{f}^2\right]\nonumber\\+ &&256\left(H^2 \dot{f}\right)^2(\ddot{f}-H \dot{f}) \bigg) \nonumber \\&& \times\frac{1}{(1-8 H \dot{f})\left[(1-8 H \dot{f}) \dot{\phi}^2+96\left(H^2 \dot{f}\right)^2\right]}\ .
\end{eqnarray}
These quantities can be expressed in terms of the background dynamical variables as $\cs = \dfrac{\mathcal{N}}{\mathcal{D}},$ with
\begin{multline}
	\mathcal{N} = H^2 \bigg(\sigma ^2 u^3 \left(z-z_c\right){}^2 \bigg(6 A \left(z-z_c\right)+ \\ z
	\left(7 \sqrt{6} x \left(z-z_c\right)+3 \lambda  y \left(z_c-z\right)-6
	x^2\right)\bigg)\\-3 u^3 z \left(z-z_c\right){}^4 \left(u-2 x^2\right)+3 \sigma
	^4 u^2 z \left(z-z_c\right){}^2 \left(1+4 x^2\right)\\
	+4 \sqrt{6}  \sigma ^6 u x z \left(z-z_c\right)+2  \sigma ^8 z\bigg)-\\u^2 z
	\overset{.}{H} \left(z-z_c\right){}^2 \bigg(3 u^2 \left(z-z_c\right){}^2 \\ +4
	\sqrt{6} \sigma ^2 u x \left(z_c-z\right)-4 \kappa ^2 \sigma ^4\bigg),\\
	\mathcal{D} = H^2 \sigma ^2 z \left(\sqrt{6} u x \left(z-z_c\right)+ \sigma ^2\right)
	\bigg(3 u^2 \left(z-z_c\right){}^2\\+2 \sqrt{6} \sigma ^2 u x
	\left(z-z_c\right)+2  \sigma ^4\bigg)\ .
\end{multline}
Similarly, the tensor sound speed becomes $\ct = \dfrac{\mathcal{N}}{\mathcal{D}}$, where 
\begin{align}
	\mathcal{N} &= 3 u z \left(z-z_c\right){}^2 \left(u \overset{.}{H}+H^2 \left(u-2
	x^2\right)\right)+3 H^2 \sigma ^2 u \bigg(2 A \nonumber \\ & \left(z_c-z\right)+z
	\left(\sqrt{6} x \left(z_c-z\right)+\lambda  y \left(z-z_c\right)+2
	x^2\right)\bigg)+H^2  \sigma ^4 z , \nonumber\\
	\mathcal{D} &= H^2 \sigma ^2 z \left(\sqrt{6} u x \left(z-z_c\right)+ \sigma ^2\right)\ .
\end{align}
To avoid ghost and gradient instabilities, the relevant quantities must satisfy
\begin{equation}
	0 \le \cs, \ct \le 1, \quad Q_{T} \equiv 1-8H \dot f >0\ ,
\end{equation}
where a negative sound square speed leads to an exponential growth of perturbation modes and consequently signals an instability.

\subsection{Cosmological parameter evolution for representative values}

In Fig.~\ref{fig:evo_best_minimal}, we plot the cosmological evolution of the fractional densities of radiation, matter, and the scalar field, as well as the effective equation of state $w_{\rm eff}$ and the deceleration parameter $q$, for the minimally coupled case, i.e., $f(\phi)=0$, with different values of the model parameter $\lambda$. The initial conditions are set deep in the  radiation-dominated regime and are chosen such that
$A_0 \gtrsim y_0$. Due to the comparable magnitudes of these two
quantities, oscillations in $w_{\rm eff}$ appear only around $N\sim0$,
after which they gradually disappear as the system approaches the
asymptotic future.

We consider $0.09\lesssim\lambda\lesssim1.5$ and observe that, during the
late-time epoch, the variable $y$ dominates over $A$. The variable $A$
attains a finite value and exhibits small oscillations around
$N\gtrsim-2.0$. Consequently, the field fractional density begins to
increase, while the field velocity $x$ also starts to oscillate. The
amplitude of the oscillations decreases as the system evolves toward the future, and for $N>1.5$, the dynamical variables approach their asymptotic values and the oscillations disappear. Since the system is free from the GB coupling in this case, $\eos$ never crosses $-1$, and the model does not exhibit any ghost or gradient instability at the perturbation level; in particular, $\ct = \cs =1$ throughout the evolution. For $\lambda=1.5$, the variable $A$ does not become negligible during the late-time epoch, unlike for the other values of $\lambda$. Thus, both components of the potential remain finite, with $y>A$, resulting in a much smaller oscillation amplitude in $\eos$ for the same set of initial conditions.
\begin{figure*}
	\resizebox{\linewidth}{!}{{\includegraphics{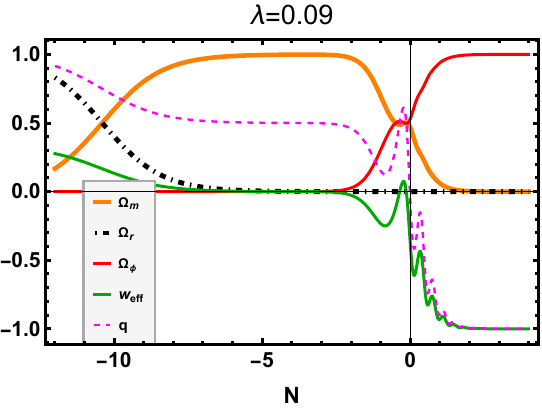},\includegraphics{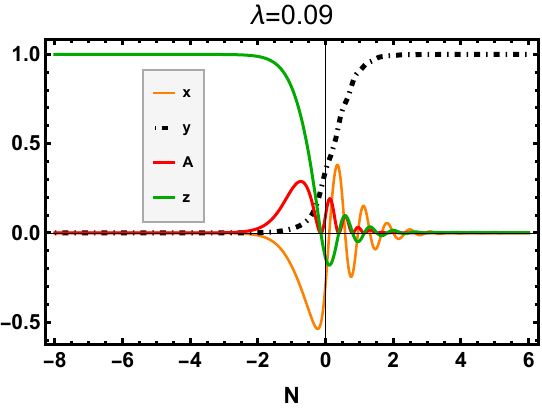}\includegraphics{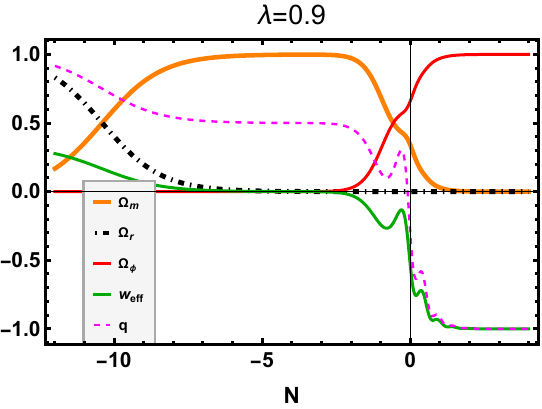},\includegraphics{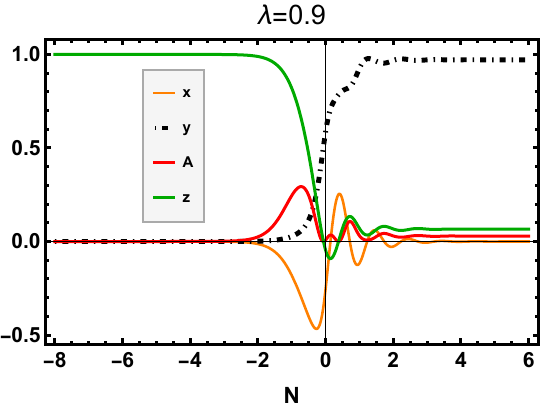}}}
	\resizebox{\linewidth}{!}{\includegraphics{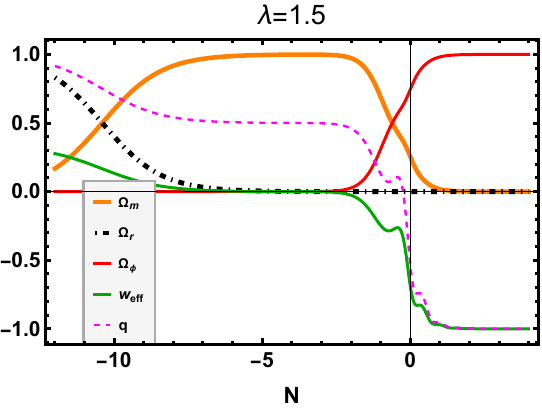}\includegraphics{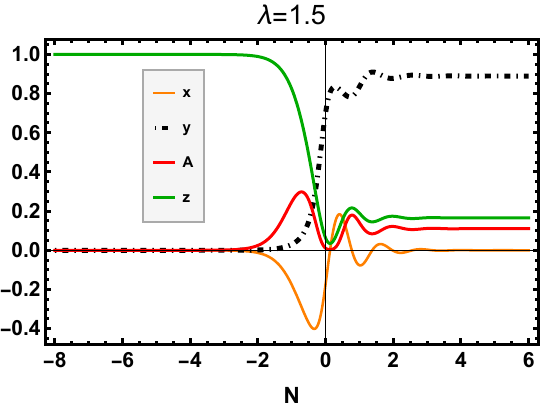}}
	\caption{Evolution of the cosmological parameters in the minimally coupled
		case, $f(\phi)=0$, plotted against $N$ for the representative
		initial values $A_0 = 10^{-29.2}$ and $y_0 = 10^{-30.4}$; the remaining initial conditions, $z_0 = 1.0$ and $x_0 = 10^{-20}$, are set deep in the radiation-dominated epoch at $N=-20$. The oscillatory behavior arises when the initial values $y_0$ and $A_0$ are comparable, so that the field settles into a minimum for a finite time before rolling again. Since the depth of the dip in the potential is controlled by $\lambda$, different values of $\lambda$ shift the oscillations toward the future.}
	\label{fig:evo_best_minimal}
\end{figure*}

\begin{figure*}
	\begin{tabular}{lc}
		(i) &\\ &\resizebox{\linewidth}{!}{\includegraphics{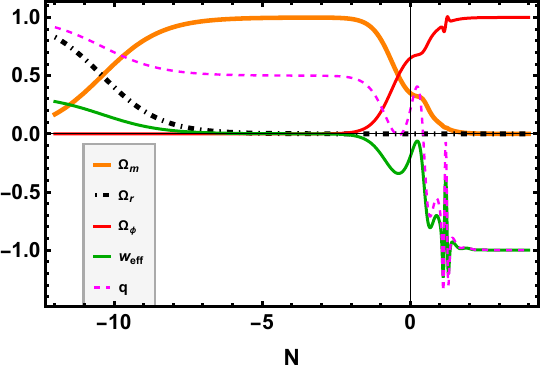}\includegraphics{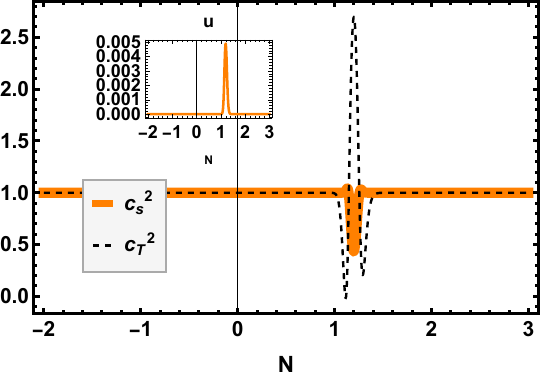} \includegraphics{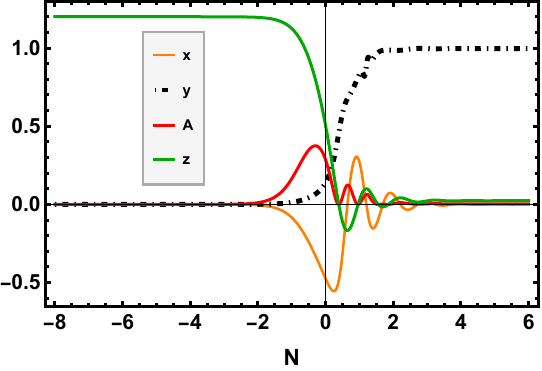}}\\
		(ii) & \\
		& \resizebox{\linewidth}{!}{\includegraphics{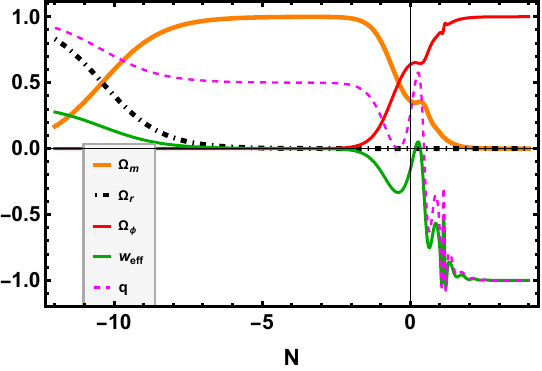}\includegraphics{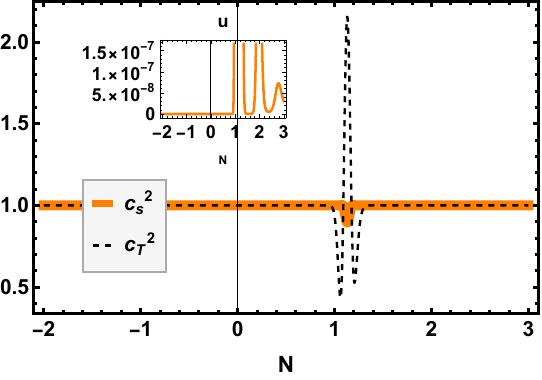} \includegraphics{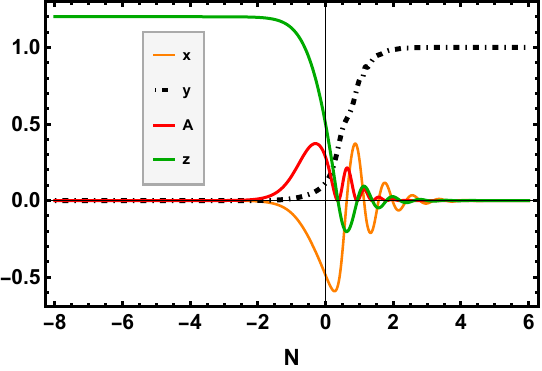}}\\
	\end{tabular}
	\caption{Evolution of the cosmological parameters and the sound speed,
		plotted as functions of $N$ for the representative values
		(i)~$\lambda = 0.31$, $A_0 = 10^{-29.56}$, $y_0 = 10^{-30.72}$,
		$z_0 = z_c = 1.2$, $x_0 = 10^{-20}$, $u_0 = 1.19$, $\sigma = 0.099$, and
		(ii)~$\lambda = 0.006$, $A_0 = 10^{-29.57}$, with the remaining initial
		conditions the same as in (i). All initial conditions are set at $N=-20$,
		deep in the radiation-dominated epoch.}
	\label{fig:evo_best_fit}
\end{figure*}

We plot the same cosmological parameters and sound-speed variables against $N$ for two distinct values, $\lambda = 0.31$ and $0.006$, in Fig.~\ref{fig:evo_best_fit}. The initial conditions $A_0 = 10^{-29.56}$, $y_0 = 10^{-30.72}$, $x_0 = 10^{-20}$, $z_0 = 1.20$, and $u_0 = 1.19$ are set deep in the radiation-dominated regime. The oscillation in $\eos$ emerges around $N\sim0$, and its amplitude decays for $N>1.5$. This oscillatory pattern differs from the minimally coupled case owing to the finite value of the non-minimal coupling function $f(\phi)$. As in the previous case, since the potential parameters are of comparable magnitude, $y$ dominates over $A$ at late times, while $A$ exhibits oscillations and takes a finite value in the range $-2<N<1.5$. All the dynamical variables nearly saturate to particular values, indicating stability of the system at the background level. However, to verify the absence of ghost and gradient instabilities, we have also plotted $\cs$ and $\ct$. Owing to the non-minimal coupling, $\eos$ crosses $-1$, and the oscillation amplitude rapidly reaches negative values in the range $-1.5<\eos<-0.5$; such large negative values cannot be achieved within the minimally coupled framework. This rapid oscillation drives $u\sim0.005$ during $N\sim1$, around which $\ct$ oscillates rapidly, with its lower amplitude approaching zero while its higher amplitude crosses into the superluminal regime $\cs>1$, whereas the scalar sound speed exhibits mild variation, remaining in the range $0.5\leq\cs<1$. In the present case, the phantom-crossing oscillation occurs at $N\sim1$; however, with a suitable adjustment of the initial conditions and model parameters, it can be shifted very close to $N\sim0$, where observational constraints apply. The rapid oscillation of $\ct$ violates the stringent bound obtained from gravitational-wave observations, $|\ct-1|<10^{-16}$ \cite{LIGOScientific:2017zic,Odintsov:2019clh,Ezquiaga:2017ekz,TerenteDiaz:2023iqk,Fier:2025}, in this case.

We find that for $\lambda=0.006$, the variation in the oscillation amplitude of $\eos$ is small, and hence $\eos$ reaches only $\sim-1.2$. A rapid oscillation occurs in $u$, but its magnitude is of order $10^{-7}$. Consequently, the minimum value attained by $\ct$ is $\sim0.5$, while $\cs$ deviates very little from $1$. Because of the small values of $u$ for both $\lambda$ values, $Q_T$ remains positive and oscillates close to $1$, indicating the absence of ghosts. The behavior of the system for these benchmark parameters shows that it can realize phantom crossing without violating the ghost condition and while satisfying the non-negative sound-speed conditions. One must note that for a very small value of $u \sim 10^{-10}$, the effect of the non-minimal coupling would be extremely small, and it can no longer drive the oscillation into the phantom regime. Thus, this demonstrates that the coupling function should be chosen such that it does not become extremely small while still allowing the model to retain observationally viable phases of the universe, since a prior study found \cite{Hussain:2025vbo} that a GB density of order $\sim 10^{-20} $ at the current epoch fails to produce phantom dynamics.

Nevertheless, in a general simulation shown in Fig.~\ref{fig:numerical_simulation_phase_space}, the model can exhibit negative scalar and tensor sound speeds, depending on the value of $u$ around $N\sim0$. If $|u|\gtrsim10^{-2}$, the model can reach more negative values, i.e., $\eos\lesssim-1.5$, which drives both sound speeds into the negative regime and causes a gradient instability. Although $Q_T$ may remain positive due to small magnitude of $u$, rendering a ghost-free system, however, it still suffers from the exponential growth of perturbation modes. Thus, we show that with a proper choice of model parameter value $\lambda, \ \sigma$, the model can avoid gradient instability. However, whether the model genuinely prefers the small-phantom regime, i.e., small values of $u$, will ultimately be determined by confronting it directly with cosmological observational datasets. In any case, the model at least exhibits an oscillatory solution near $N\sim0$ and eventually settles into a stable de Sitter solution.

\section{Conclusion}
\label{sec:conclusion}

In this paper, we have investigated an oscillatory dark-energy scenario in the Einstein-scalar-Gauss--Bonnet framework, in which the effective equation of state can cross the phantom divide multiple times before settling into a stable de Sitter future. The model combines a scalar-field potential containing exponential and quadratic terms with a Gaussian GB coupling function localized around the minimum of the potential, such that the field is temporarily trapped, undergoes oscillations, and eventually evolves toward the asymptotic attractor. The oscillatory behavior depends on the depth of the potential, which can be regulated through the choice of initial conditions in the deep radiation-dominated phase. When the exponential and quadratic contributions are of comparable magnitude, the resulting oscillations in the effective equation of state remain confined to the low-redshift regime and do not attain excessively large amplitudes.

We have established a dynamical-systems framework to identify the regions of the initial-condition and model-parameter space for which the system produces a stable late-time de Sitter solution, with oscillations occurring during the dark matter--dark energy transition epoch. The model admits an extended radiation- and matter-dominated era, followed by a scalar-field-dominated late-time attractor that can exhibit phantom crossing. We find that, in the majority of cases where the GB coupling function is of order $\mathcal O{(-2)}$, the effective equation of state of the system reaches very large negative values, consequently driving both the scalar and tensor sound square speeds into the negative regime and resulting in gradient instability. We find that for sufficiently small values of the exponential potential parameter $\lambda$, close to $\sim 10^{-3}$, together with suitable initial conditions, the effective equation of state can cross the phantom divide while remaining close to $-1.2$. In this regime, both the tensor and scalar sound speeds remain positive, thereby avoiding gradient instability. We also find that, irrespective of whether the sound speeds become negative or remain positive, the system does not exhibit a ghost instability because the magnitude of the GB coupling remains sufficiently small.

The model therefore provides a dynamical realization of transient oscillatory dark energy that can be consistent at both the background and perturbation levels. However, whether the theoretically stable region of the parameter space is also favored by current low-redshift observations, and what distinct perturbation signatures the oscillatory crossings may leave behind, requires confronting the model directly with cosmological observational data. Additionally, the transient oscillatory phenomenon can be explored in the context of interacting dark matter--dark energy frameworks, where the induced oscillations during the current epoch may have non-trivial observational consequences that can be constrained using large-scale structure observations.

\appendix

\section{Center Manifold Theorem}
\label{app:cmt}
When all eigenvalues have non-zero real parts, the linearized system provides a clear characterization of the local stability. However, when one or more eigenvalues have zero real parts while the remaining eigenvalues have negative real parts, the linear stability analysis becomes inconclusive. Center manifold theory \cite{B_hmer_2012,Bahamonde:2017ize}
provides a framework to analyze such cases by reducing the dynamics near
the equilibrium to an invariant manifold tangent to the subspace associated
with the zero-real-part eigenvalues.

Consider an autonomous dynamical system
\begin{equation}
	\mathbf{X}'=\mathbf{F}(\mathbf{X}),\qquad \mathbf{X}\in\mathbb{R}^{n},
\end{equation}
where a prime denotes differentiation with respect to the evolution
variable. Let \(\mathbf{X}_*\) be an equilibrium point satisfying
\begin{equation}
	\mathbf{F}(\mathbf{X}_*)=0.
\end{equation}
Introducing the perturbation \(\mathbf{U}=\mathbf{X}-\mathbf{X}_*\), the
system can be expanded about the equilibrium as
\begin{equation}
	\mathbf{U}'=\rm J \ \mathbf{U}+\mathbf{G}(\mathbf{U}),
\end{equation}
where
\begin{equation}
\rm 	J=D \ \mathbf{F}(\mathbf{X}_*)
\end{equation}
is the Jacobian matrix, $\rm D$ is derivative operator, evaluated at the equilibrium point and
\begin{equation}
	\mathbf{G}(\mathbf{U})=\mathcal{O}(|\mathbf{U}|^2)
\end{equation}
contains the nonlinear terms. Since $\rm J$ is an \(n\times n\) matrix, it
has \(n\) eigenvalues, which can be divided into three categories:
(i) stable (\(s\)), corresponding to eigenvalues with negative real parts,
(ii) unstable (\(u\)), corresponding to eigenvalues with positive real
parts, and (iii) center (\(c\)), corresponding to eigenvalues with zero real parts. The space \(\mathbb{R}^{n}\) is then decomposed as
\begin{equation}
	\mathbb{R}^{n}=E^{s}\oplus E^{u}\oplus E^{c},
\end{equation}
where \(E^{s}\), \(E^{u}\), and \(E^{c}\) denote the subspaces spanned by
the stable, unstable, and center eigenvectors, respectively. If the unstable subspace is empty, the system can be transformed locally
into the form
\begin{equation}
	\begin{aligned}
		\mathbf{x}' &= A\mathbf{x}+\mathbf{f}(\mathbf{x},\mathbf{y}),\\
		\mathbf{y}' &= B\mathbf{y}+\mathbf{g}(\mathbf{x},\mathbf{y}),
	\end{aligned}
\end{equation}
where $\mathbf{x}\in\mathbb{R}^{c}$ and $\mathbf{y}\in\mathbb{R}^{s}$, with
$c$ and $s$ denoting the dimensions of the center and stable subspaces,
respectively. The matrix $A$ has eigenvalues with zero real parts, while
all eigenvalues of $B$ have negative real parts. The nonlinear functions satisfy
\begin{equation}
	\mathbf{f}(\mathbf{0},\mathbf{0})=\mathbf{0},
	\qquad
	\nabla\mathbf{f}(\mathbf{0},\mathbf{0})=\mathbf{0},
\end{equation}
and
\begin{equation}
	\mathbf{g}(\mathbf{0},\mathbf{0})=\mathbf{0},
	\qquad
	\nabla\mathbf{g}(\mathbf{0},\mathbf{0})=\mathbf{0}.
\end{equation}
A center manifold is a geometrical space that can be locally represented as
\begin{equation}
	W^c(\mathbf{0})=
	\left\{
	(\mathbf{x},\mathbf{y})\in\mathbb{R}^{c}\times\mathbb{R}^{s}
	\mid
	\mathbf{y}=h(\mathbf{x}),\ |\mathbf{x}|<\delta
	\right\},
\end{equation}
where $h(\mathbf{0})=\mathbf{0}$ and $\nabla h(\mathbf{0})=\mathbf{0}$ for
sufficiently small $\delta$. These conditions imply that
$W^c(\mathbf{0})$ is tangent to the center eigen-space at the critical
point.

Since a trajectory that starts on the center manifold remains on it, the
relation $\mathbf{y}=h(\mathbf{x})$ must hold along the trajectory.
Differentiating this relation with respect to the evolution variable gives
$\mathbf{y}'={\rm D} h(\mathbf{x})\mathbf{x}'$. Using
$\mathbf{y}'=B\mathbf{y}+\mathbf{g}(\mathbf{x},\mathbf{y})$ and setting
$\mathbf{y}=h(\mathbf{x})$, we obtain the invariance equation
\begin{equation}
	{\rm D}  h(\mathbf{x})\left[A\mathbf{x}+\mathbf{f}\left(\mathbf{x},h(\mathbf{x})\right)\right]
	=
	Bh(\mathbf{x})+\mathbf{g}\left(\mathbf{x},h(\mathbf{x})\right).
\end{equation}

As an illustrative example, consider the two-dimensional autonomous system
\begin{equation}
	\begin{aligned}
		x'&=-x,\\
		y'&=0.
	\end{aligned}
\end{equation}
The equilibrium condition requires
\begin{equation}
	x=0,
\end{equation}
while \(y\) remains arbitrary. The system therefore has a one-dimensional
family of equilibrium points,
\begin{equation}
	\mathcal{M}_{\mathrm{eq}}=\left\{(0,y):y\in\mathbb{R}\right\}.
\end{equation}
The Jacobian is
\begin{equation}
\rm 	J=
	\begin{pmatrix}
		-1 & 0\\
		0 & 0
	\end{pmatrix},
\end{equation}
with eigenvalues $-1$ and $0$, so that the $x$ direction is the stable one
and the $y$ direction is the center one. Since the center subspace is the
$y$-axis, the center manifold can be written as
\begin{equation}
	W^c=\left\{(x,y):x=h(y)\right\},
\end{equation}
where, as stated above, the definition of the center manifold requires
\begin{equation}
	h(0)=0,\qquad h'(0)=0.
\end{equation}
The relation $x=h(y)$ must remain satisfied along a trajectory on the
center manifold. Differentiating with respect to the evolution variable, we obtain
\begin{equation}
	x'=h_{,y}(y)\,y'.
\end{equation}
On the manifold, $x'=-x=-h(y)$, and hence
\begin{equation}
	h_{,y}\,y'=-h(y).
\end{equation}
Since $y'=0$, it follows that
\begin{equation}
	h(y)=0.
\end{equation}
On the center manifold, $x=0$, and the reduced dynamics is therefore
\begin{equation}
	y'=0.
\end{equation}
Thus, perturbations along the center direction neither grow nor decay,
while perturbations in the stable $x$ direction decay according to
$x'=-x$. The equilibrium manifold is therefore locally attracting in the
transverse direction but neutrally stable along itself.

Similarly, in our case, the critical point $P_1$ contains two free
parameters, which we denote by $z$ and $u$. The critical points therefore
form a two-dimensional equilibrium manifold, $\mathcal{M}_{1}$. If the
Jacobian evaluated at $P_1$ possesses two eigenvalues with zero real parts,
the corresponding center subspace is two-dimensional. The relation between
the center subspace and the equilibrium manifold can be established by
considering the tangent vectors
\begin{equation}
	\frac{\partial P_1}{\partial z},
	\qquad
	\frac{\partial P_1}{\partial u}.
\end{equation}
Since every point on $\mathcal{M}_{1}$ is an equilibrium, it satisfies
\begin{equation}
	\mathbf{F}\left(P_1(z,u)\right)=\mathbf{0}.
\end{equation}
Differentiating this relation with respect to $z$ and $u$ gives
\begin{equation}
	{\rm J} (P_1)\frac{\partial P_1}{\partial z}=\mathbf{0},
	\qquad
{\rm J} (P_1)\frac{\partial P_1}{\partial u}=\mathbf{0}.
\end{equation}
Thus, both tangent vectors to the equilibrium manifold belong to the kernel
of the Jacobian and hence correspond to zero-eigenvalue directions. Since
the two tangent vectors are linearly independent, they span a
two-dimensional subspace. As the center subspace is also two-dimensional,
we obtain
\begin{equation}
	T_{P_1}\mathcal{M}_{1}=E^c_{P_1}.
\end{equation}
Consequently, the equilibrium manifold $\mathcal{M}_{1}$ itself serves as
the local center manifold associated with $P_1$. For $P_1$, the four
nonzero eigenvalues have negative real parts, indicating that perturbations
transverse to the equilibrium manifold decay with the evolution of the
system. The two zero eigenvalues correspond to the tangent directions of
the equilibrium manifold and therefore lead to neither growth nor decay of
perturbations along the manifold. A trajectory starting sufficiently close
to the equilibrium manifold is consequently attracted toward the manifold,
but not necessarily toward a particular equilibrium point on it. Thus,
$P_1$ represents a transversely attracting equilibrium manifold, with
neutral dynamics along its tangent directions.

For the critical point $P_2$, the variable $z$ remains arbitrary, and hence
the critical points form a one-dimensional equilibrium manifold,
\begin{equation}
	\mathcal{M}_2=\left\{P_2(z)\right\}.
\end{equation}
Since every point on this manifold is an equilibrium, we have
\begin{equation}
	\mathbf{F}\left(P_2(z)\right)=\mathbf{0}.
\end{equation}
Differentiating with respect to the free parameter $z$ gives
\begin{equation}
	{\rm J} (P_2)\frac{\partial P_2}{\partial z}=\mathbf{0}.
\end{equation}
Thus, the tangent direction to the equilibrium manifold corresponds to a
zero eigenvalue of the Jacobian. When the Jacobian has exactly one
eigenvalue with zero real part, the center subspace is one-dimensional, and
therefore
\begin{equation}
	T_{P_2}\mathcal{M}_2=E^c_{P_2}.
\end{equation}
Hence, the one-dimensional equilibrium manifold $\mathcal{M}_2$ itself
serves as the local center manifold associated with $P_2$. The stability of
$P_2$ is then determined by the eigenvalues transverse to this equilibrium
manifold. Thus, $P_2$ is also transversely attracting and neutrally stable
along the equilibrium manifold.

\bibliographystyle{apsrev4-2}

\bibliography{egb}

\end{document}